\documentclass{article}

\usepackage{xurl}
\usepackage{arxiv}
\usepackage{float}
\usepackage[utf8]{inputenc} % allow utf-8 input
\usepackage[T1]{fontenc}    % use 8-bit T1 fonts
\usepackage{hyperref}       % hyperlinks
\usepackage{url}            % simple URL typesetting
\usepackage{booktabs}       % professional-quality tables
\usepackage{amsfonts}       % blackboard math symbols
\usepackage{nicefrac}       % compact symbols for 1/2, etc.
\usepackage{microtype}      % microtypography
\usepackage{lipsum}		% Can be removed after putting your text content
\usepackage{graphicx}
\usepackage{natbib}
\usepackage{doi}
\usepackage{multicol}
\usepackage{balance}
\usepackage{array}
\usepackage{multirow}
\usepackage{amsmath}
\usepackage{amssymb}
\usepackage{listings}
\usepackage{xcolor}

\definecolor{codegreen}{rgb}{0,0.6,0}
\definecolor{codegray}{rgb}{0.5,0.5,0.5}
\definecolor{codepurple}{rgb}{0.58,0,0.82}
\definecolor{backcolour}{rgb}{0.95,0.95,0.92}

\lstdefinelanguage{json}{
    basicstyle=\small\ttfamily,
    showstringspaces=false,
    breaklines=true,
    backgroundcolor=\color{backcolour},
    commentstyle=\color{codegreen},
    keywordstyle=\color{magenta},
    stringstyle=\color{codepurple},
}

\title{AcademiaOS: Automating Grounded Theory Development in Qualitative Research with Large Language Models}

\author{ {\hspace{1mm}Thomas Übellacker}\thanks{Maastricht University; based on\\ original bachelor thesis.\\Live version: \href{https://academia-os.org}{academia-os.org}\\\href{https://github.com/thomasuebi/academia-os}{github.com/thomasuebi/academia-os}}
	\texttt{} 
	} 
\renewcommand{\shorttitle}{}

\hypersetup{
pdftitle={AcademiaOS: Automating Grounded Theory Development in Qualitative Research with Large Language Models},
pdfauthor={Thomas.~Übellacker},
pdfkeywords={Large Language Models, Software Design, Information Retrieval, Research Automation, Theory Development},
}

\begin{document}
\maketitle
\begin{multicols}{2}
    \begin{abstract}
        AcademiaOS is a first attempt to automate grounded theory development in qualitative research with large language models. Using recent large language models' language understanding, generation, and reasoning capabilities, AcademiaOS codes curated qualitative raw data such as interview transcripts and develops themes and dimensions to further develop a grounded theoretical model, affording novel insights. A user study (n=19) suggests that the system finds acceptance in the academic community and exhibits the potential to augment humans in qualitative research. AcademiaOS has been made open-source for others to build upon and adapt to their use cases.
    \end{abstract}

    % keywords can be removed
    \keywords{“Large Language Models”, “Software Design”, “Information Retrieval”, “Research Automation”, “Theory Development”}

    \section{Introduction}
    Qualitative researchers are confronted with tedious and costly language tasks when making sense of interview or focus group transcriptions, reports, policies, field notes from observational research, diaries, case studies, and other sources. Transcribing and coding a typical interview takes several hours (\cite{pope2000qualitative}; \cite{barrett2018data} p. 63). A transcript or set of codes is often just the beginning, as the researcher tries to make sense of the raw data and transform it into something useful to gain insight and knowledge and develop theories that may describe patterns and phenomena.

    Qualitative data may be collected through various tools, while purposive sampling directs the research (\cite{chun2019grounded}, p. 3). Interviews are a direct and easy way of exploring phenomena, from open, unstructured formats that allow researchers to dive deeper into studied participants' narratives and experiences to more rigidly structured formats, essentially questionnaires, that provide more control to the researcher. Interviews are usually recorded and transcribed for further processing (\cite{barrett2018data} p. 63). A more relaxed way of studying subjects' views is focus groups, where the researcher speaks to a small group of study participants to guide the conversation and debate about a given topic and to create rich data in the form of notes or transcription (\cite{kitzinger1995qualitative} p. 301). Another qualitative data collection methodology, prevalent primarily in ethnography, is field observation, where the researcher observes study subjects in a natural setting (\cite{emerson2001participant}, p. 352). (\cite{emerson2001participant}, pp. 356 - 358) describe generating data from ethnographic qualitative research as a process, starting with a researcher's mental notes from immersing in the situation, followed by writing down jotted notes, to use as support for later generated, more extensive field notes. Ethnographic research also often builds on diaries and written notes, while organizational research might build on case studies, reports, policies, and other secondary sources. The bandwidth of qualitative data sources is broad, but most are textual or have established ways to transform them into textual data (i.e., transcribing).

    Researchers often follow established coding practices to manage large volumes of unstructured textual sources. Coding often involves systematically generating a codebook (\cite{weston2001analyzing}) to code transcriptions. Another popular approach is the "Gioia method" (\cite{gioia2013seeking}), where the researcher extracts emerging patterns and concepts from the source documents directly to aggregate and interpret them only in the following steps. These developed codes can then be further used for qualitative data analysis and theory development. The notion of developing theoretical models from data is called grounded theory development (\cite{chun2019grounded}).

    \cite{gioia2013seeking} define a transparent process for analyzing qualitative data to develop theoretical models from interviews. Their process is meant to move a researcher from raw qualitative data to increasingly abstract conceptual categories, starting with initial coding, where a researcher codes and underlines relevant concepts in the data, resulting in an extensive list of first-order concepts still stated in the language of the source document. They then use these first-order concepts to produce a list of more abstract second-order themes that try to encapsulate the idea of the first-order codes in more academic language. Lastly, they aggregate the second-order themes into even more abstract "aggregate dimensions." These are then used as the basis for developing theories. \cite{gioia2013seeking} mention the importance of understanding the dynamic relationships between these concepts but have yet to offer a concrete methodology to get to these. They assume that by following this "Gioia method," a researcher has familiarized themselves well enough with the underlying literature to understand these relationships.

    Another approach to grounded theory development is the \cite{eisenhardt1989building} method, which focuses on constructing models from case studies. This approach starts with a within-case analysis to gain familiarity with the data and generate a preliminary theory. From there starts a highly iterative process, which \cite{eisenhardt1989building} calls "Shaping Hypotheses," iteratively comparing data and developed constructs and verifying that the emergent relationships between the developed constructs fit with the evidence in the data. They treat case studies as replications of experiments, either strengthening or weakening a hypothesis.

    Structured approaches to qualitative analysis and theory development, such as those introduced by \cite{gioia2013seeking} and \cite{eisenhardt1989building}, were criticized for being used as mere "templates," often oversimplifying or ignoring the underlying methodological assumptions to fit in cases where those assumptions do not hold (\cite{gioia2022curated}, p. 241) or that using these templates limit methodological innovation in qualitative research (\cite{gioia2022curated}, p. 242). Even Gioia emphasizes that the methodology named after him should not be understood as a "cookbook" but as a tool adapted to each research context. He highlights the importance of researchers accounting for informants' interpretations and constructions of reality in their reporting and not merely applying the methodology as a rigid framework (\cite{gioia2022curated}, p. 235). Building on the existing literature, it is clear that structured approaches like those of \cite{gioia2013seeking} and \cite{eisenhardt1989building} provide a framework for grounded theory development. However, they also highlight the necessity for flexibility and adaptation in qualitative research. Qualitative research tasks, both data collection and analysis, are time-consuming and expensive and limit the empirical data that a single research team can study. \cite{kindsiko2019poor} support this notion by outlining that sample size in empirical research depends on funding and the size of the research team. However, \cite{bowen2008naturalistic} outlines how the sample size influences research validity and suggests limiting sample size by saturation, i.e., stopping the addition of more interviews, case studies, or other samples when those do not add significant new information. Now, how do we add research rigor by increasing sample size while keeping manual effort low? The answer might be found in computational automation.

    In quantitative research, automating data preparation and theory development is an actively researched topic under names such as "data mining" or "machine learning," where computer programs learn from observations to develop mathematical models, allowing them to estimate future instances in a positivist paradigm. Qualitative research problems, however, come with less structured or codifiable information and rely on the knowledge and interpretation of the researcher. At the same time, with the rise of large language models (LLMs), we have access to a technology platform, shifting the paradigm of computational understanding and reasoning on textual data to a level close to humans, paired with extensive general knowledge. This new technology platform presents an opportunity to simulate well-defined research processes at scale. Tasks such as coding 100 interview transcripts are time-consuming for a single researcher. The hypothesis is that with the proper setup, LLMs can process all transcripts in parallel within a few minutes. A researcher in the field of organization theory might consider using two or three case studies to develop a theoretical model. When finding empirical evidence in case studies is an automated, parallelizable process, the barrier to using 20 - 30 different case studies is drastically minimized, offering opportunities for more statistically relevant qualitative studies.

    Thus, the potential for automating parts of the qualitative research process with LLMs is ripe for exploration. This paper explores the following research question: "How can a foundational open-source platform be effectively designed and implemented to leverage large language models to automate grounded theory development?". To that end, AcademiaOS is proposed and implemented as an open-source platform for automating or augmenting grounded theory development tasks such as coding, dimension aggregation, and theory development. AcademiaOS offers the scientific community a new methodology for conducting qualitative research that is transparent, accessible, and extensible (through its open-source character) and allows for broader evidence through the cost-effectiveness of parallelizing analysis of multiple qualitative sources simultaneously. This system could have a far-reaching impact in the social sciences, particularly in organization theory, but also in other disciplines where qualitative data is relevant.

    The rest of this paper describes AcademiaOS and the technology behind it in more detail. In addition, an exploratory survey has been conducted to learn how students, professionals, and researchers use AcademiaOS and to guide future development. The platform is made open-source on GitHub under the MIT license, and a running instance is hosted on academia-os.org.

    \begin{figure*}
        \centering
        \includegraphics[width=0.8\linewidth]{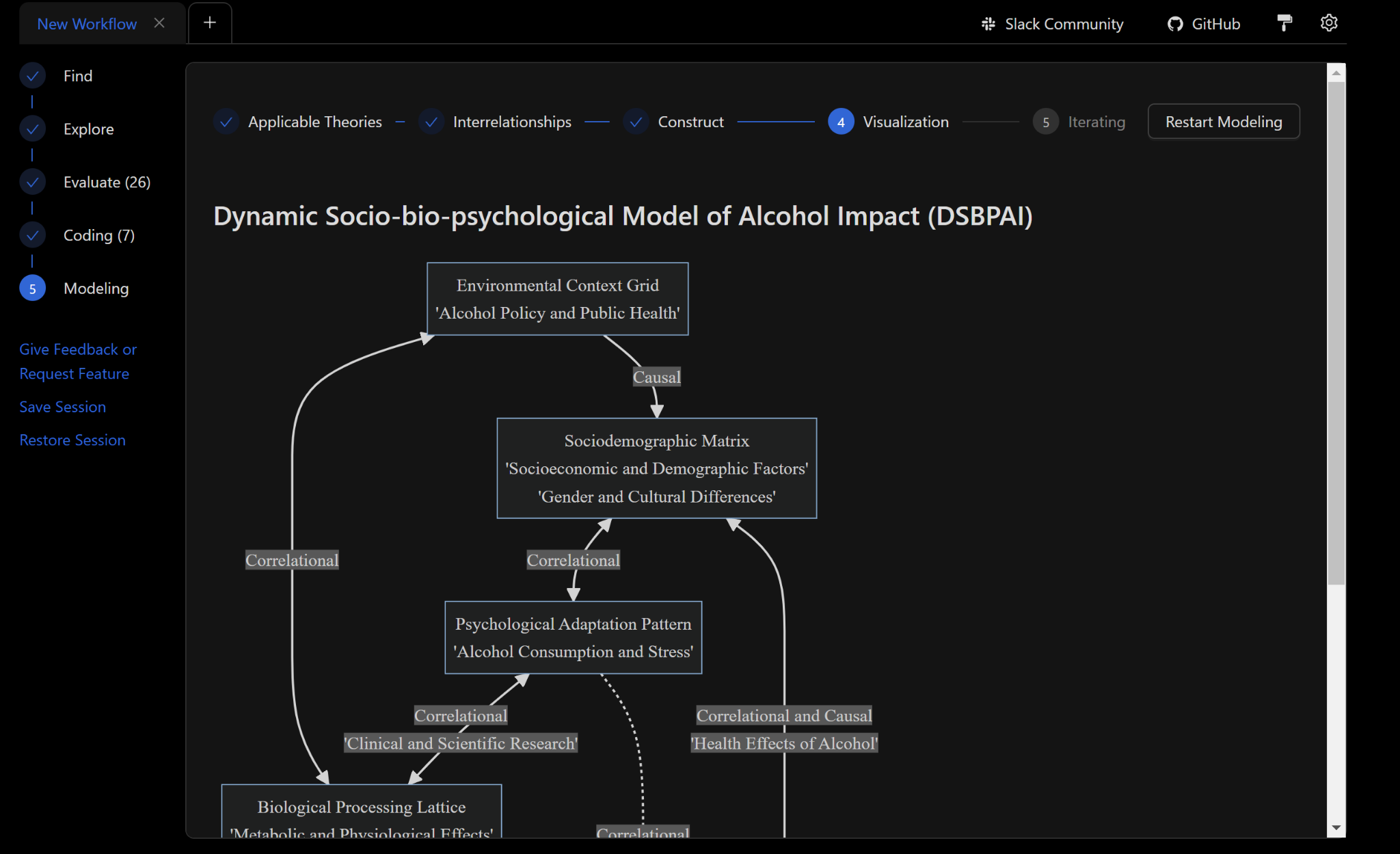}
        \caption{AcademiaOS user-interface}
        \label{fig:enter-label}
    \end{figure*}

    \section{Earlier Work}
    There have been attempts to automate qualitative analysis. \cite{berente2019research} make an effort to develop a theoretical process for computationally intensive grounded theory development, proposing a theoretical, computational methodology to automate the following four steps for grounded theory development: (1) Sampling and data collection, (2) synchronic analysis, (3) lexical framing and (4) diachronic analysis using artificial-intelligence-based tools. They describe the computational process as evolving around a predefined but dynamic lexicon rather than synchronously "coding" emerging concepts. They propose to mine concepts using taxonomies.

    \cite{marathe2018semi} discuss the possibility of automating the coding of interviews based on predefined human-annotated codebooks. Similar approaches have been implemented by \cite{lennon2021developing}, according to their own analyses resulting in human-level accuracy. \cite{rietz2021cody} propose a semi-automated supervised machine-learning solution that learns coding rules from human annotators and applies them to a broader dataset. Still, defining rules or annotating a codebook beforehand assumes the researcher already has assumptions about the patterns in the analyzed qualitative literature. Furthermore, the academic papers mentioned above used old machine-learning techniques and ignored LLMs, while some commercial platforms are slowly adopting them. The commercial platform \cite{ATLAS.ti} announced test versions of their automated coding features in early 2023, splitting qualitative literature into paragraphs and coding them one by one using OpenAI's LLMs. Other commercial platforms like elicit.org emerged, primarily using LLMs to automate literature review processes. Still, more is needed to know about how researchers use these new AI features on those platforms. Also, these applications automate only small portions of qualitative research processes and have yet to dive into the field of automating grounded theory development. This leads to the question of whether grounded theory development can be automated with LLMs.

    \section{Large Language Models}
    Large language models (LLMs) are a new technology platform based on transformer models, pre-trained on large datasets using self-supervision, encoding general and transferable knowledge in billions of parameters \cite{roberts2020much}. Unexpected capabilities such as reasoning and interaction emerged from learning "simple" next-token prediction at scale. These pre-trained foundation models are often fine-tuned to follow instructions (\cite{ouyang2022training}), to return structured output, or to be conversational (as seen with ChatGPT). While older models such as BERT would generally be regarded as LLM, in this paper, the term is exclusively used for models with a performance similar to or better than the GPT-3 base models. LLMs have been popularized and mass-adopted with the recent release of ChatGPT at the end of 2022. They have been applied to automate processes throughout industries (\cite{wulf2023towards}, p. 4).

    An inference interaction with an LLM typically consists of a natural language prompt (input) and completion (response). In OpenAI's conversational models used throughout this paper (GPT-3.5 and newer), the inference prompt may consist of multiple "messages": generic system messages that set the frame and a history of user and assistant messages (see Appendices 1-11 for examples).

    Without the need for fine-tuning, LLMs exhibit the ability to learn and generalize from information in the inference prompt (\cite{brown2020language}). One-shot or few-shot learning refers to passing samples in the prompt, whereas zero-shot learning refers to providing no samples but having the model complete an explicit instruction. This type of inference contrasts with common task-specific fine-tuning and is commonly called "in-context learning" (\cite{dong2022survey}).

    Although being pre-trained, LLMs store tremendous amounts of implicit knowledge in their parameters, that knowledge remains limited in its depth and currency and requires costly training to update. Therefore, augmenting the LLM inference with an information retrieval system, an architecture commonly called "Retrieval-Augmented Generation" (RAG), has been shown to reduce hallucination and improve factualness and interpretability (\cite{lewis2020retrieval}). As general models, LLMs can measure semantic similarity between two text strings. Their similarity can be calculated computationally efficiently using cosine similarity on their LLM-internal vector representation. RAG uses this straightforward way of information retrieval to concatenate retrieved, relevant text for in-context learning (\cite{lewis2020retrieval}). Augmenting LLM inference with information retrieval from raw input documents may help automate theory development.

    \section{Approach}
    To explore the current possibilities of automating grounded theory development with LLMs, I propose, develop, and test a piece of software that orchestrates LLM inferences with human supervision. AcademiaOS is a platform for qualitative researchers to automate their grounded theory development. The platform guides the user through a predefined process, and while most of the data analysis and theory development parts are automated, the user owns the supervision and control. To ensure user privacy and high maintainability, for potential future developers and open-source contributors to not worry about frontend-backend interaction, most of the computation is performed locally in the browser, consuming external APIs such as the OpenAI developer platform for LLM inference directly.
\end{multicols}
\begin{figure}[!ht]
    \centering
    \includegraphics[width=0.8\textwidth]{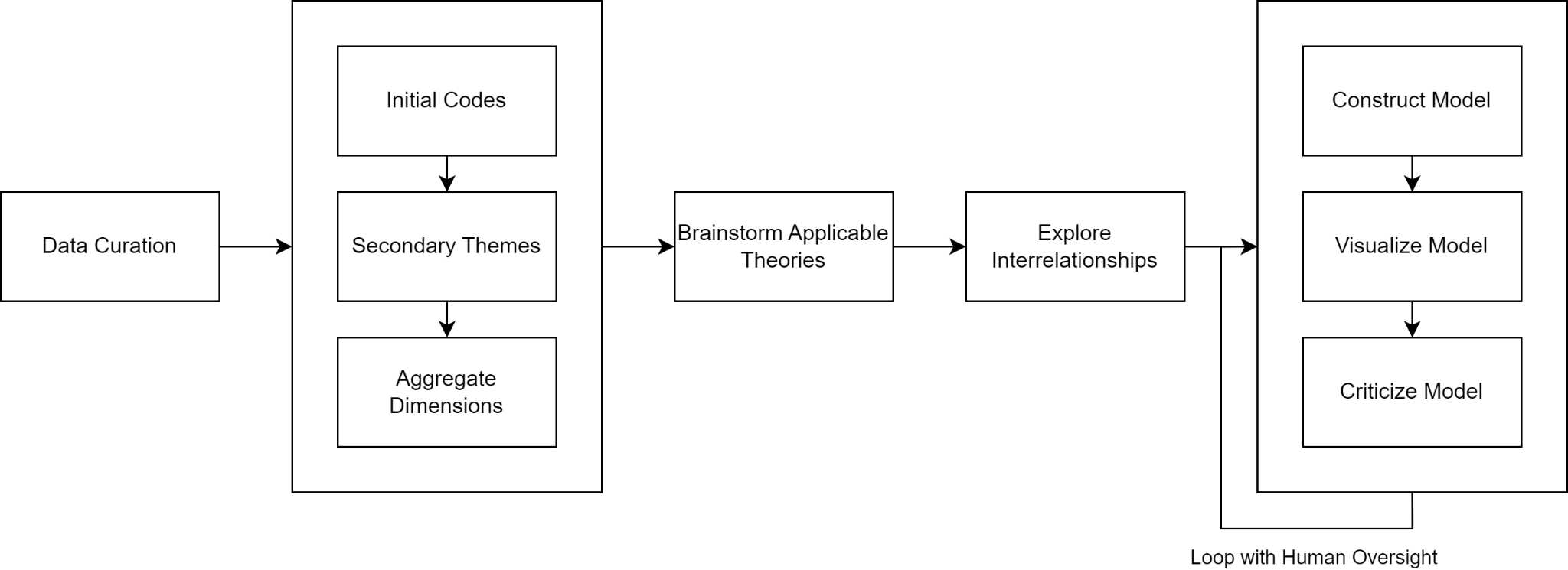}
    \caption{High-level process for grounded theory development with AcademiaOS}
    \label{fig:Fig 2}
\end{figure}
\begin{multicols}{2}
    \subsection{Data Curation}

    \vspace{-2mm}
    Qualitative research is a broad field working with various qualitative information sources, from interviews over policies to organizational case studies. To ensure versatility, the platform affords users to curate their source documents in multiple ways. If the user already has a corpus of curated sources, they can directly upload one or many source documents. The system extracts textual information for further use directly from documents in PDF, JSON, or TXT format. To support scanned PDF documents, the system pre-processes problematic PDF files by applying optical character recognition via Adobe PDF Services. Every file is transformed into a text string and handled as an individual source document.

    A second option helps the user curate relevant academic literature from free-text search. The search term is used to retrieve academic literature from the SemanticScholar search engine that indexes more than 200 million academic papers. To interface with the SemanticScholar REST API, SemanticScholarJS was introduced as an open-source TypeScript library and published under the MIT license. SemanticScholar returns at most 100 papers, which are then vector-embedded using OpenAI's text embeddings and re-ranked according to their abstracts' semantic similarity to the initial search query and correspondingly re-ranked. Any papers without an abstract are filtered out, resulting in a few documents relevant to the user's search query.

    \subsection{Data Coding}
    Assuming that the user might have an idea about their research's direction, the platform affords the user to add a free-text remark or choose from an LLM-generated research question, which is used for in-context learning in the following inference prompts. This interactivity respects the expertise of the researcher while providing AI-assisted efficiency.

    \vspace{-2mm}
    \begin{table}[H]
        \centering
        \begin{tabular}{|>{\raggedright\arraybackslash}p{60pt}|>{\raggedright\arraybackslash}p{66pt}|>{\raggedright\arraybackslash}p{60pt}|}
            \hline
            \textbf{Level}       & \textbf{Description}                                                                                                   & \textbf{Example}                                                   \\
            \hline
            Initial codes        & Short text strings describing emergent themes, concepts, and patterns in the language of the raw document.             & “Mentions of renewable energy”, “References to cost-saving”        \\
            \hline
            Second-order themes  & Aggregate and interpret semantically similar initial codes, expressed in more abstract, academically correct language. & “Environmental Sustainability Trends”, “Efficiency in Operations”  \\
            \hline
            Aggregate dimensions & More abstract, quantifiable concepts derived from second-order themes.                                                 & “Corporate Environmental Responsibility”, “Operational Excellence” \\
            \hline
        \end{tabular}
        \vspace{1mm}
        \caption{Coding hierarchy (\cite{gioia2013seeking})}
        \label{tab:Tab 1}
    \end{table}

    \vspace{-8mm}
    Following the Gioia method (\cite{gioia2013seeking}), the system initializes a three-step data analysis process. Implementing the Gioia method aligns the platform with established qualitative research practices, lending academic rigor to the process. This structured approach facilitates the systematic transformation of raw data into initial codes, second-order themes, and aggregate dimensions, used later for grounded theory development.

    First, the system creates initial codes from the raw documents. Initial codes are short text strings describing emergent themes, concepts, and patterns in the language of the raw document. To that end, every input document is split into chunks of max. 10,000 characters to overcome context-size limitations and process arbitrarily large input. A chunk overlap of 50 characters was introduced to mitigate losing information in the intersection of two chunks. Every chunk is then inserted as context in the LLM inference prompt with a natural language instruction to return a JSON array. The output is then parsed and stored in a global array of initial codes.

    Next, the initial codes are used to develop second-order themes. Second-order themes aggregate and interpret semantically similar initial codes and are expressed in a more abstract, academically correct language. To that end, the initial codes are split into reasonably-sized chunks and used in-context with a natural language instruction to return a JSON object of second-order themes as keys and arrays of related initial codes as values. This JSON object is again parsed and stored as a global object.

    Finally, the second-order themes are used to develop aggregate dimensions. These aggregate dimensions are even more abstract, quantifiable concepts. Since second-order themes already limit the number of codes to a manageable count, further splitting into chunks is not necessary and the second-order themes are directly used in-context together with a natural language instruction to return a JSON object of aggregate dimensions as keys and arrays of related second-order themes as values. This JSON object is again parsed and stored as a global object for further use.
    \subsection{Theory Development}
    \begin{table}[H]
        \centering
        \begin{tabular}{|>{\raggedright\arraybackslash}p{60pt}|>{\raggedright\arraybackslash}p{66pt}|>{\raggedright\arraybackslash}p{60pt}|}
            \hline
            \textbf{Stage}         & \textbf{Description}                                             & \textbf{Example}                                  \\
            \hline
            Applicable theories    & Brainstorming applicable existing theories to scrutinize against & “Triple Bottom Line”                              \\
            \hline
            Concept tuples         & Developing tuples of potentially interrelated concepts           & “Green Energy, Cost Reduction”                    \\
            \hline
            Relationship Summaries & Generating grounded concept tuple relationship summaries         & “Energy Efficiency Link”                          \\
            \hline
            Theoretical Model      & Generating the textual representation of the model               & “Sustainable Operational Model”                   \\
            \hline
            Visualization          & Generating the visual representation of the model                & “Sustainable Operational Model Graph”             \\
            \hline
            Critique               & Critizising the developed model for further iteration            & “Missing link to long-term sustainability impact” \\
            \hline
        \end{tabular}
        \vspace{1mm}
        \caption{Steps in theory development}
        \label{tab:Tab 2}
    \end{table}
    \vspace{-5mm}

    In the next step, the developed initial codes, second-order themes, and aggregate dimensions are used to develop grounded theories. (\cite{wacker1998definition}, pp. 364-367) found that to be considered 'good,' a theory must exhibit the following virtues: uniqueness, conservatism, generalizability, fecundity, simplicity, internal consistency, empirical riskiness, and abstraction. To ensure some of those virtues, such as uniqueness, conservatism, and fecundity, the aggregate dimensions are used for the machine to "brainstorm" applicable theories that exist in the literature. As this brainstormed output is not intended to return accurate information but merely to scrutinize the obvious and increase serendipity, this step happens in a single inference prompt without information retrieval.

    Next, to ground any interrelationships of concepts used in the final model in the data, the LLM first creates concept tuples, then uses these concept tuples for retrieval-augmented generation, searching for relevant paragraphs (chunks of max. 1000 characters, 50-character chunk-overlap) in the raw documents using cosine similarity search on their OpenAI vector embeddings and the concept tuples. Next, given the four most similar paragraphs, the LLM is prompted to generate a one-sentence summary about the relationship of a given concept tuple. This summary might include information about any evidence for correlation or causation, directedness, strength, mediatedness, conditional interaction, dynamics, or feedback loops.

    Previously developed applicable theories, interrelationships, aggregate dimensions, second-order codes, and the user's free-hand remarks are then fed into an LLM prompt, instructed to describe a comprehensive, novel theoretical model.

    Next, another one-shot LLM inference transforms a given model description into a MermaidJS visualization. MermaidJS is a scripting language for creating flowcharts and similar graphics. Converting model descriptions into visualizations using MermaidJS not only aids in understanding complex theories but also provides a straightforward, communicative tool for sharing findings. First-order and second-order codes provided in the prompt are used to visualize evidence and support the notion of groundedness.

    Finally, a last zero-shot LLM inference criticizes the developed model for novelty, conciseness, clarity, and theoretical insight. The user can then decide whether to start another modeling iteration with the critique fed into the modeling prompt, resulting in a feedback loop with human oversight.

    \subsection{Notes on the Technical Implementation}
    The system was initially considered a conglomerate of separate frontend and backend services. While this design decision would have come with the advantage of controlling any requests that go to the OpenAI developer platform, it would have slowed down development, maintenance, and hosting for the course of the project and beyond for future developers and open-source contributors. Thus, the decision was to run the application entirely in the browser. TypeScript as a primary programming language was chosen for its strong typing compared to vanilla JavaScript and, thus, higher maintainability and more robust code quality. Additionally, TypeScript's extensive web development community offers abundant resources, libraries, and support, which is crucial for both the initial development and future iterations of the platform. Moreover, the decision to run the entire application in the browser, using ReactJS as a frontend framework and AntDesign as a component library, was driven by the need for an accessible, responsive, user-friendly interface. The ReactJS framework, through its component-based architecture, allows for efficient rendering and state management, which is essential in a dynamic environment where the state can frequently change through user interactions. The browser-based approach also aligns with the privacy concerns inherent in qualitative research. AcademiaOS minimizes the risks associated with data transfer and storage on remote servers by processing data locally on the user's machine.

    All results reported in this paper are based on OpenAI’s gpt-4-1106-preview model. To increase traceability and transparency, AcademiaOS affords users to easily set up a proxy to Helicone, an open-source monitoring platform specifically for LLM applications. Through Helicone, users maintain complete control and insight over their OpenAI API usage and the corresponding costs. In addition, Helicone logs all LLM inference requests. Monitoring is convenient during development to see where the LLM goes wrong and tune prompts. In addition, the platform affords saving and restoring developed codes and models for further development using JSON serialization and deserialization. This comes in especially handy during the development and testing phases to mitigate performing redundant steps and to facilitate a smooth development and testing workflow. The serialized data can also easily be shared between multiple users.

    The field of large language models is actively researched and worked on. During this project, OpenAI and other vendors released new models with new capabilities. OpenAI's most recent model, GPT-4-turbo, allows for larger context windows in a single inference prompt but also comes with an optional "JSON-mode" parameter, which ensures that the inference response is in a valid, parsable JSON format. Since many of AcademiaOS' inference prompts expect data in a structured format (e.g., arrays of values), enforcing the JSON mode significantly increased the application's robustness. Still, as LLM inference can occasionally yield invalid or unexpected results, rerunning the system on the same input often results in varied outputs, thereby increasing the probability of obtaining a correct response with subsequent attempts.

    \subsection{Notes on Epistemology and Ontology}

    Grounded theory development, as practiced by \cite{gioia2013seeking} or \cite{eisenhardt1989building}, generally follows a constructivist epistemology (\cite{charmaz2021pursuit}). Constructivism in grounded theory proposes that knowledge is actively constructed through the interaction between the researcher and the researched subject. It argues for a subjective nature of reality and knowledge generation where the researcher's perceptions shape the interpretative process (\cite{levers2013philosophical}, pp. 3-4). Further, grounded theory assumes that multiple realities of subjective constructions of individuals exist in a fluid nature of the reality of social interactions. As grounded theory development generates theories from observations of empirical data without previously established theories, grounded theory development is an inherently inductive approach.

    \vspace{-2mm}
    \begin{table}[H]
        \centering
        \begin{tabular}{|>{\raggedright\arraybackslash}p{60pt}|>{\raggedright\arraybackslash}p{66pt}|>{\raggedright\arraybackslash}p{60pt}|}
            \hline
            \textbf{Aspect}   & \textbf{Grounded Theory Approach}                                         & \textbf{Impact of LLMs}                                     \\
            \hline
            Epistemology      & Constructivist: knowledge is subjectively constructed.                    & Adds objectivity in data processing.                        \\
            \hline
            Ontology          & Multiple subjective realities in social interactions.                     & Introduces a digital reality, offering new perspectives.    \\
            \hline
            Methodology       & Inductive; theories from empirical data without pre-established theories. & \multirow{5}{*}{-}                                          \\
            \hline
            Researcher's Role & Interpretation and contextualization of data.                             & Continues interpreting LLM output.                          \\
            \hline
            LLMs' Role        & \multirow{5}{*}{-}                                                        & Process data objectively, support/challenge interpretation. \\
            \hline
        \end{tabular}
        \vspace{1mm}
        \caption{Epistemology and Ontology}
        \label{tab:Tab 3}
    \end{table}

    \vspace{-6mm}
    Automating grounded theory development with large language models challenges epistemological and ontological assumptions. Epistemologically, Introducing LLMs adds an element of objectivity to the traditionally subjective approach. They process qualitative data in a way independent of a researcher's biases. However, the researcher keeps the responsibility of interpreting and contextualizing the LLM's output, preserving the constructivist nature of the process. At the same time, LLMs ontologically add a digital reality to the existing subjective realities. This digital reality offers an algorithmically created perspective on the processed qualitative data that can support or challenge the researcher's interpretation.

    \subsection{Notes on Temperature, Robustness and Serendipity}

    To get an intuition on how model temperature changes the system's output, outputs of gradual uniform temperature increase are recorded and qualitatively analyzed. To accommodate deterministic reproducibility, a seed parameter (324234) has been utilized with the employed gpt-4-1106-preview model for grounded theory development from a single podcast transcript from the How I Built This podcast.

    Appendices 61 - 65 show that the three attempts with temperatures of 0, 0.5, and 1 resulted in consistent and robust outputs. In contrast, the attempt with a temperature of 1.5 became unstable, resulting in non-sensical text in the latter part of the model description and an invalid MermaidJS visualization. The attempts with temperatures of 0 and 0.5 resulted in internally consistent and integrated models. The attempt with a temperature of 1 produced a theoretical model with less integrity, meaning not or only loosely connected aspects and thus missing a consistent narrative. An increased temperature could, in this case, not be observed to lead to a higher degree of serendipity, that is, the quality of pleasant yet unexpected results. This might highlight the still objective, reproducible, and deterministic nature of grounded theory development and that randomness in the data analysis does not necessarily lead to better results. However, while providing a sense of intuition, the sample size of this experiment (n=1) might need to be larger to generalize.

    \section{Survey}
    An exploratory, qualitative survey has been conducted to assess user interaction with AcademiaOS and to guide future development. Researchers, professionals, and students with a background in qualitative research were chosen via convenience sampling for this evaluation. The participants (n=19), who were deliberately selected for their relevance to the platform's intended user base and thus thought to yield the most informative and relevant data for assessing AcademiaOS, were provided with a link to a survey hosted on the Qualtrics platform.

    The study's methodology involved guiding participants through the AcademiaOS platform. This was achieved by direct interaction with the platform, through a video demonstration, or a combination, depending on the participant's preference. Eight respondents reported testing the platform with a corpus of university policies on ChatGPT use that was provided alongside the survey. Other participants reported using datasets such as interview transcripts or academic papers from a basic literature review. Only 4 participants based their answers solely on the video demonstration (see Appendix 17). Following their experience with the platform, participants were asked to share their insights and opinions. The survey primarily consisted of open-ended questions, allowing for detailed and nuanced feedback.

    The feedback encompasses various aspects of the platform, including user interface design, assessing the user's initial interaction, feature exploration, and usability, but also the effectiveness of the system's output and perceived impact on the user's immediate research environment. Additional exploratory questions were asked about their current qualitative research processes and context.

    Partial survey responses have been recorded. However, the 28 responses have been filtered and cleaned only to include 19 finished responses for further analysis. The other responses were very sparse, and it could not be trusted that they spent sufficient time familiarizing themselves with the platform. Another participant stated, "I don't generally work with qualitative data," whose response was also filtered due to not belonging to the target population. Of the 19 residual responses, 14 were students, and the remaining participants were professionals and academic researchers (see Appendix 12). This might skew the responses to students' perspectives. \cite{peterson2014convenience} argue that convenience samples of students are problematic as they do not generalize to the general consumer population. However, they also explain that students might be an appropriate sample if they represent a population of interest. As the sampled students have at least some levels of experience in qualitative research, they sufficiently represent the studied group.

\end{multicols}

\section{Results}

\begin{table}[H]
    \centering
    \begin{tabular}{|>{\raggedright\arraybackslash}p{100pt}|>{\raggedright\arraybackslash}p{165pt}|>{\raggedright\arraybackslash}p{165pt}|}
        \hline
        \textbf{Aspect}                 & \textbf{Key Findings}                                                                                                           & \textbf{Implications/Concerns}                                                                                      \\
        \hline
        Current Research Methodology    & Diverse primary and secondary sources used; adoption of various data collection and analysis methods, including NLP techniques. & Evolving research methodologies blending qualitative and quantitative elements; need for advanced analytical tools. \\
        \hline
        AI Tools in Research            & Varied use of AI tools like ChatGPT, PyTorch for tasks like brainstorming and coding; concerns over reliability.                & Importance of AI in research, coupled with a need for accuracy and reliability in AI tools.                         \\
        \hline
        Initial Interaction/Exploration & Mixed ease of use; suggestions for improved UI and guidance; challenges with coding and theory development features.            & Need for more intuitive UI and comprehensive user guidance.                                                         \\
        \hline
        Usability and Satisfaction      & General satisfaction with coding process; challenges with multilingual documents and content variation.                         & Importance of improving document retrieval and coding for diverse content.                                          \\
        \hline
        Theory Development              & Satisfaction with theory development but concerns over complexity and relevance to research questions.                          & Need for simpler, more targeted models for theory development.                                                      \\
        \hline
        Impact on Research              & Positive impact on research efficiency; concerns over ethical implications, quality, biases, and AI replacing humans.           & Balance between AI utility and ethical considerations; addressing quality and bias issues.                          \\
        \hline
        Future Use and Recommendations  & Excitement about AcademiaOS’ future; willingness to continue using and recommending the platform.                               & Potential for wider adoption and ongoing development of the platform.                                               \\
        \hline
    \end{tabular}
    \vspace{1mm}
    \caption{Key Findings}
    \label{tab:Tab 4}
\end{table}
\begin{multicols}{2}
    \subsection{Status Quo}

    \vspace{-4mm}
    Participants were asked to reflect on their current qualitative research methodology. They reported using primary sources such as interviews, observations, surveys, and panels and secondary data sources such as case studies, reports, meta-analyses, historical data, reports, and expert opinions (Appendix 13).

    Asked about their research methods, they reported data collection tools such as voice recordings, Microsoft Forms, or Google Forms for online surveys, data analysis methods such as thematic analysis, content analysis, literature review and case law analysis, empirical research, and knowledge discovery from existing datasets. One respondent mentioned testing qualitative data with quantitative research tools, while other respondents mentioned statistics and natural language processing techniques (such as named-entity recognition and topic modeling), hinting that there are tools to "quantify" qualitative data and that in research settings, the line between qualitative and quantitative research is often blurred. Interestingly, one respondent even mentioned using language models and the previously mentioned RAG architecture for their qualitative research (see Appendix 14). For systematic theory development, respondents mentioned grounded theory methodology, conceptual framework development, mind maps, and flowcharts, while some respondents see themselves merely as model users and less as creators or are unsure about their methodologies (see Appendix 15).

    Inquiring about the AI tools they utilize in their research (see Appendix 16), some participants reported limited to no use of AI tools, not having used them or only using them rarely. Many respondents mentioned using ChatGPT with varying degrees of reliance and purposes such as brainstorming, idea generation, speeding up coding processes, or narrowing down searches. One respondent even noted the need for extensive fact-checking due to high error rates with ChatGPT. Other participants mentioned more specific AI and machine learning platforms and libraries: Fairseq, HuggingFace, Pytorch, OpenAI, AzureML, TensorFlow, scikit-learn, RapidMiner, KNIME, NLTK, SpaCy, Gensim and BERT. These reported AI tools underline the perceived importance of quantifying qualitative data through natural language processing.

    \subsection{Initial Interaction and Feature Exploration}
    The second part of the survey aimed to learn how participants perceived the platform's initial interaction and feature exploration. The participants generally rated the platform as "somewhat easy" to learn, with some variance (see Appendix 18). Participants mentioned that the video helped them sufficiently understand how the system works. One participant suggested adding popups to guide the user, and another suggested that the affordances were not entirely clear: "[...] you should mention that the user can also upload non-pdf files [...]" (appendix 19). Overall, participants reported finding it easy to find and use the option to search for academic papers (see Appendix 20) and to upload their own sources (see Appendix 21). However, they had more issues finding and using the options to "code" their sources (see Appendix 22) and to develop the theory (see Appendix 23). In general, participants found the user interface to be "somewhat" to "strongly" intuitive and user-friendly, with only one outlier who "strongly disagrees" (see Appendix 24).

    As for difficulties navigating the platform (see Appendix 25), participants reported they expected more descriptive prompts and guidance: "I couldn't find the coding help function," "the insert key thing was rather hidden," and "about the code part, didn't really understand how to use it." One user noted that some buttons needed to be visible on mobile and experienced long loading times. Some improvement suggestions (see appendix 26) included "make it [...] less technical", to add "filters for finding papers", to "add examples" and an "instructions page" to better guide users. Furthermore, better error handling with retry policies was suggested, together with the idea to "combine different searches or datasets in one coding tab." These feedback points are actionable and specific enough to guide the platform's development.

    \subsection{Usability, Effectiveness, and User Satisfaction}
    Study participants seem to be "somewhat" to "extremely" satisfied with the coding process, with only one "extremely dissatisfied" outlier (see Appendix 27). Even more promisingly, participants think the system would create better codes than when they code manually (see Appendix 28). Only one participant thinks the system-generated output codes would be "somewhat worse" than their own. This is to be taken with a grain of salt as most respondents will have answered that only intuitively, without any actual comparison. Challenges in coding lie in multilingual documents (see Appendix 29). One respondent noted that when the source documents were not purposefully selected and had too much variation in their content, the machine could not generate meaningful codes: "[...] it was not able to catch that and make codes relevant to the research question". This seems to be an issue related to the paper retrieval and less so with the coding process. The average participant generally rated AcademiaOS as "very effective" in augmenting the coding process on qualitative data (see Appendix 30).

    Regarding the theory development process, participants reported being somewhat to extremely satisfied (see Appendix 31). The respondents also reported that the system-produced output was perceived as better than potential manual development, with only three respondents each reporting that the system would do "somewhat worse" or "about the same" than when done manually (see Appendix 32). When asked about instances where the produced theory did not meet their expectations (see Appendix 33), one participant mentioned, "It was kind of complex, and I often didn't really get where it was going with the theory.", indicating that they would have preferred a more straightforward model, while another one wrote "[...] theory seems to cover the content of the sources pretty generally but is not really about the research question I posed." indicating that there was a mismatch between the free-hand remarks and the model output. Despite these instances, respondents perceived the platform to be moderately effective to extremely effective, with a skew to the latter (see Appendix 34).

    \subsection{Participant-Reported Impact}
    When asked about whether AcademiaOS would impact their qualitative research process, most participants answered with "probably yes" to "definitely yes" (Appendix 35), mentioning reasons such as speeding up their research processes ("Much faster," "It will speed up the research," "It would make the coding and theory production much faster"), to act as a tool for inspiration ("Multiple draft iterations to inspire/shortcut the manual process," "I would use it to rapidly prototype theories [...]", "[...] compare and maybe find things I missed before."), and as general research support ("It will urge many scientists to increase their throughput [...] potential to mitigate the information retrieval and retention of humans", "[...] it would help me to link concepts easier", "Making it easier for me to do my research, especially in hours where I cannot concentrate that much") (see Appendix 36). When asked about how AcademiaOS would change the output of their academic research (see Appendix 37), participants noted "more varied and tested theories, better theories through more input ideas, more feedback," "use the models created as part of a clear visualization" and "[...] to be more reproducible". All respondents agree that AcademiaOS helps them ground their research on more data (see Appendix 41) and that it helps them conduct research faster (see Appendix 42). At the same time, most participants also agree that it helps them find more impactful and unexpected results (see Appendix 43).

    However, participants also noted certain limitations such as ethical considerations ("It feels like cheating to me… I would not feel comfortable using it.", "It would make me feel too replaceable"), doubts about quality ("I also don't quite trust the quality (mostly from experience with ChatGPT), I would always want to check it for what it misses.", "[...] it might fail to generate very latent codes") and doubts of potential (reinforced) biases ("[...] it may cause people to not properly look at the nuances of their interviews or papers", "Currently it is using a single AI model, which makes it difficult to rule out its biases.") (see Appendix 38). In addition, some participants fear being replaced by AI (see Appendix 40).

    Respondents generally feel excited about the new possibilities that AcademiaOS provides (see Appendix 39). Most participants even stated they would continue using the system for their research (see Appendix 45). To an even greater extent, they would recommend AcademiaOS to their friends (see Appendix 46), indicating that they might know potential users who fit the intended target group better. Participants generally also feel excited about future versions of AcademiaOS (see Appendix 47), pointing out that despite the discussed doubts and limitations, the technology is promising for their research field.

    \section{Example Applications}
    This section presents a series of illustrative use-cases for AcademiaOS.

    \subsection{Modeling University Policies on ChatGPT Use}

    A corpus of 46 scraped public university policies on ChatGPT and LLMs were uploaded to AcademiaOS to build a theoretical model to answer the research question: "How can educational institutions adapt their academic integrity policies to effectively address the challenges posed by AI writing tools like ChatGPT?". From the 46 papers, AcademiaOS produced 1019 first-order codes, which in the next step were reduced and interpreted to 67 second-order themes and, in turn, reduced to 7 aggregate dimensions (see Appendix 49). After brainstorming applicable existing theories (see Appendix 50), the system then investigated the interrelationships between concept tuples (see Appendix 51) to generate the first iteration of a theoretical model (see Appendix 52). The system then critiqued the model (see Appendix 53) to produce the second iteration of the theoretical model (see Appendix 54). Note that the user is in control of the number of iterations.

    The result was the Integrated Education Adaptation Framework for AI Readiness (IEAFAR) as a comprehensive model for educational adaptation to artificial intelligence. Its six dimensions, AI Integration and Educational Dynamics (AI-ED), Adaptive Compliance and Policy Evolution (ACPE), Academic and Instructional Resilience (AIR), AI-Driven Educational Innovation (ADEI), Ethical, Legal, and Societal Harmony (ELSH), and Collaborative Stakeholder Engagement (CSE) emphasize the importance of holistic, iterative processes in adapting to AI, with interconnected relationships among these dimensions. IEAFAR promotes stakeholder collaboration, focusing on continuous learning, flexible adaptation, and balancing innovation and ethical, legal, and societal considerations.

    \subsection{Exploring the “How I Built This” Transcript Dataset}
    The "How I Built This" podcast is a format interviewing innovators about the things (such as products or companies) they have created. The hypothesis is that AcademiaOS can explore the rich and extensive interview dataset to detect interesting patterns in how entrepreneurs start projects. The system coded the transcripts of 4 podcast episodes to 499 first-order codes, 35 second-order themes, and 7 aggregate dimensions (see Appendix 55). It then generated applicable theories and explored the interrelationships between concept tuples before generating the first iteration of the theoretical model. This time, the AI-generated the "Dynamic Entrepreneurial Synthesis Model (DESM)" (see appendix 56) grounded in the podcast transcripts.

    The Dynamic Entrepreneurial Synthesis Model describes the multifaceted journey of business development, integrating personal growth with the development of innovative products. Community, market understanding, social responsibility, and strategic financial planning are central. Further, the model suggests that strong relationships, market traction, and effective leadership are core to successful entrepreneurship.

    \subsection{Modeling the Cardiovascular Health Effects of Alcohol}
    The following example is slightly different, as policies or interview transcripts are not uploaded. Instead, it starts with a free-text search for academic papers to further explore, code, and model the topic. AcademiaOS selected 26 academic publications from SemanticScholar to use for further processing. It then codes the abstracts of these publications to 395 first-order codes (see appendix 57), 35 second-order themes, and 7 aggregate dimensions. These codes are then used to create the first-iteration model (see Appendix 58), which gets critiqued (see Appendix 59) to create the second-iteration model (see Appendix 60). In this case, the system was able to create an easy-to-grasp model called "The Dynamic Socio-bio-psychological Model of Alcohol Impact (DSBPAI)," connecting the "environmental context grid" with the "individual health trajectory," facilitating new research angles. The model shows reciprocal feedback mechanisms and causal correlations, highlighting the multidimensionality of influences from molecular to societal scales.

    \section{Discussion}
    The following section critically examines the findings from the implementation and the user study of AcademiaOS. This discussion aims to contextualize these findings within the broader scope of existing research, address implications, acknowledge limitations, and suggest directions for future research. The primary objective of this study was to design and implement an open-source platform to automate or augment grounded theory development. An accompanying user survey was designed to assess the platform's effectiveness. The findings suggest that the platform is well-received within the academic community. AcademiaOS' capabilities in typically labor-intensive tasks such as coding qualitative data and developing theoretical models constitute a significant step in augmenting human researchers in qualitative research.

    The user survey provides relevant insight into the target population's opinion and experience with AcademiaOS. Respondent's diverse use of primary and secondary qualitative sources and various data collection and analysis methods emphasize the multifaceted nature of qualitative research. The inclusion of AI tools such as ChatGPT and other natural language processing technologies highlights the trend toward integrating quantitative methods into qualitative research. This blurs the boundaries between traditional qualitative research and quantitative research. By leveraging technologies like AcademiaOS, researchers can uncover patterns and insights that might otherwise remain obscured due to the voluminous nature of qualitative data. This new technology could shift how qualitative research is conducted, making it more efficient and comprehensive while maintaining and following the established framework and corresponding research paradigms of grounded theory development (\cite{gioia2013seeking}; \cite{eisenhardt1989building}). AcademiaOS embodies a pragmatic approach, joining the subjectivity of human interpretation with the efficiency and scale of LLMs.

    The general satisfaction with the coding process and the belief that the system could generate better codes than manual coding is promising. However, participants also noted challenges in coding multilingual documents and dealing with varied content sources, suggesting that there is room for improvement in the system's sensitivity to other data types. Also, the participant's satisfaction with the theory development process and illustrated example applications such as policy or interview analysis appear promising. However, user feedback suggests that the system might still have issues aligning closely with a user's research goals.

    The reported potential impact of AcademiaOS speeding up research processes and providing inspiration is a significant positive outcome. However, ethical concerns and doubts about quality and bias indicate the need for a robust framework ensuring ethical use and quality control. The participant's excitement about the technology's potential and willingness to recommend and continue using the platform signals a positive outlook for AcademiaOS in the academic community.

    \section{Limitations}
    While AcademiaOS introduces a novel approach to automating grounded theory development, this work has several limitations. First, through its reliance on LLMs, the system inherits some of the common limitations those models exhibit. \cite{chen2023beyond} found that LLMs perform worse than common information retrieval systems in answering questions factually correctly, especially in a few-shot in-context learning context. The theory development use-case, however, is not per se a knowledge generation task that requires high levels of accuracy or informativeness, as long as those metrics are present enough together with coherency, relevancy, helpfulness, and validity, all measures where \cite{chen2023beyond} found good performance. As LLMs' output is open, sometimes unexpected out-of-range completions, such as incorrect MermaidJS visualization script syntax or wrong JSON fields, can occur. This can only partly be mitigated by writing stricter prompts (e.g., specifying the output schema or providing a specific example). \cite{kocon2023chatgpt} found that state-of-the-art AI solutions generally outperform current LLMs on common NLP tasks, suggesting that specialized models could also perform better on some functions where AcademiaOS employed LLMs, such as the coding process. A benchmarking of different technologies was not in the scope of this work. The inherent bias of models like GPT-4 (\cite{bubeck2023sparks}, pp. 86 - 89) might pose challenges to sensitive topics, such as when working with protected attributes. However, since the idea of grounded theory development is to base any hypotheses on the curated source data, there is little room for biased interpretation, neither for human nor machine inference.

    While the designed LLM-supported process allows for coding qualitative documents for grounded theory development at scale, documents such as interview transcripts and their antecedent codes might miss latent information and nuances that a researcher who conducted the interviews themselves might possess, immersing themselves more profoundly in the studied environment. An LLM can only partially compensate for that with the extensive general knowledge it draws on, potentially leading to misinterpretations or overgeneralizations in the developed theories. Therefore, qualitative research might develop to be a joint effort between human and machine inference. \cite{jiang2021supporting} studied human-AI interaction in qualitative research. They pointed out another possible limitation: researchers might be reluctant to have AI to eliminate the "uncertainty" from their research efforts. They argue that researchers value the inefficiencies in working with qualitative data and that mistakes in interview coding, for instance, lead to higher serendipity and new perspectives. Automation might be a hindrance in this process. However, survey participants reported expecting AcademiaOS to help them gain more unexpected results (see Appendix 43). \cite{bouschery2023augmenting} explore the same aspect in product innovation teams whose methodologies are similar to those of academic researchers and found that those teams profit from larger problem spaces and solution spaces when working jointly with AI.

    There are also data privacy concerns. AcademiaOS, in its current form, draws on the OpenAI developer platform for LLM inferences. Therefore, sensitive data not meant to be shared with external entities can not be processed with the proposed system for ethical and legal reasons. However, to ensure data privacy, the system can be modified to run on a self-hosted LLM (such as a Llama2 or a Mistral 7B instance), ensuring complete data control.

    Another limitation is scope. AcademiaOS is primarily designed for qualitative research in fields such as organization theory and may not be as effective for other research areas where qualitative data plays a secondary role or where grounded theory development is not an appropriate methodology. Also, AcademiaOS is far from automating the entire qualitative research process that researchers should overly rely on. Instead, AcademiaOS is a tool supporting the researcher in understanding the data and modeling interesting patterns that might guide or inspire their qualitative research.

    As for methodological limitations, the user survey was designed to be exploratory and might not give a representative image of AcademiaOS' performance in the field. While the study gave informative and extensive subjective insight, the participant sample was too small to reach data saturation, and insights were given on initial impressions rather than on actual work using this new tool. Further research might be necessary to understand the technology's impact more accurately.

    \section{Ethics, Risks, and Trustworthy AI}

    What follows is a critical reflection on the potential impact and systemic risks associated with using the tool in qualitative research. Autonomous systems are often criticized or feared for not aligning with human values. This is a complex problem, as it is hard to identify a correct set of values the AI should align with, living in a world of different moral stances, leaving the option to identify a common ground of values that would hold in a pluralist reality (\cite{gabriel2020artificial}, pp. 14-16). This might also draw on political discourse on how to aggregate different stances (see, e.g., \cite{rawls2017theory}) or which universal values to agree on (e.g., the Universal Declaration of Human Rights). The \cite{european2019ethics} made an attempt to translate liberal European values into concrete guidelines for AI systems.

    \cite{o2017weapons} introduces the concept of "Weapons of Math Destruction," arguing that algorithms, automation, and big data potentially lead to undesirable outcomes. She explores multiple effects that might make autonomous systems dangerous. Firstly, the author argues that while seemingly objective, these systems can amplify or perpetuate bias in the data they are fed. This might be problematic in fields where a biased system can lead to unfair outcomes. Secondly, O'Neil criticizes opaque algorithms and lack of transparency, which can lead to accountability issues. Thirdly, she emphasizes that algorithms, unlike human agents, can work at scale, affecting thousands of people at once. Lastly, she criticizes potentially destructive feedback loops.

    I will argue that AcademiaOS is designed around trustworthy AI standards that aim to mitigate these kinds of destructive effects. \cite{thiebes2021trustworthy} identify five foundational principles of trustworthy AI that are largely compatible with established guidelines from public institutions, such as the Ethics Guidelines by the \cite{european2019ethics}: Beneficence, non-maleficence, autonomy, justice, and explicability. In the following, I will analyze AcademiaOS using these five principles.

    Firstly, beneficence is defined as the quality of promoting the well-being of humans and respecting basic human rights (\cite{thiebes2021trustworthy}). \cite{european2019ethics} further specifies these as "fundamental rights," that is, respect for human dignity, freedom of the individual, respect for democracy, justice and the rule of law, equality, non-discrimination and solidarity, and citizens' rights. AcademiaOS implicitly promotes the well-being of humans by promoting scientific discovery and, therefore, the advancement of knowledge and education. This helps researchers conduct rigorous research more efficiently and promotes broader access to qualitative research methods with minimal violation of individuals' fundamental rights. According to \cite{thiebes2021trustworthy}, these qualities promote trusting beliefs such as benevolence, helpfulness, and purpose, adding to the notion of trustworthy AI.

    Secondly, non-maleficence is the quality of preventing harm to people (\cite{thiebes2021trustworthy}. That entails the protection of human dignity and mental and physical integrity. Galactica from Meta is a recent case study for malicious science communication with LLMs. Galactica is an LLM trained on scientific literature to produce corresponding content. It was shut down only three days after being opened to the public for being criticized for hallucinating false information while articulating it in a scientifically convincing way (\cite{heaven2022meta}). This might have led to severe consequences, such as polluted journals with made-up "knowledge." \cite{boiko2023autonomous} published another example of LLM-powered scientific discovery, specifically in chemical research. They outlined the risk of dual-use application, allowing non-scientists to synthesize dangerous and illegal substances by a simple prompt to the autonomous research agent. One could argue that AcademiaOS is prone to be misused in a scenario similar to the Galactica case, where an AI system autonomously publishes papers at scale and pollutes academic databases with false information. However, AcademiaOS is not an agent autonomously interacting with the world and creating harm on its own. In its current form, it is a mere tool controlled by a human researcher at every step. While AcademiaOS might lower the entry barrier to grounded theory development and thus also for potential misuse, through the still high computational cost, the necessary human component, and therefore time spent generating theoretical models with AcademiaOS, this is not an issue of scalable impact as \cite{o2017weapons} would have warned about. Furthermore, even with malevolent misuse, e.g., through the spreading of wrong information, the accountability question does not change to when done manually.

    \begin{table}[H]
        \centering
        \begin{tabular}{|>{\raggedright\arraybackslash}p{50pt}|>{\raggedright\arraybackslash}p{150pt}|}
            \hline
            \textbf{ Principle} & \textbf{Description in AcademiaOS Context}                                                                  \\
            \hline
            Beneficence         & Promotes well-being and respects basic human rights by advancing scientific knowledge and education.        \\
            \hline
            Non-maleficence     & Prevents harm through a controlled, human-guided approach, minimizing misuse and biased outcomes.           \\
            \hline
            Autonomy            & Ensures human agency and control in the research process, keeping a human-in-the-loop approach.             \\
            \hline
            Justice             & Addresses potential biases and promotes fairness through real-time moderation and open-source transparency. \\
            \hline
            Explicability       & Offers transparency and traceability through open-source nature and integration with Helicone.              \\
            \hline
        \end{tabular}
        \vspace{1mm}
        \caption{Trustworthy AI principles}
        \label{tab:tab 5}
    \end{table}

    \vspace{-4mm}
    Other risks for harm, as pointed out by \cite{o2017weapons}, are reflected or amplified bias and potential perpetuation or destructive feedback loops. To illustrate this, consider a hypothetical scenario set in the realm of social sciences research in the 1950s. An AI system is employed to analyze social data. The AI, constrained by the prevailing societal norms and limited data of its time, could inadvertently reinforce existing biases. It might generate theories that perpetuate gender or racial stereotypes because the data of that era is heavily skewed. This could have significant ethical implications and potentially end in a destructive feedback loop. I would argue that this naïve empiricism is inherent in grounded theory development – whether automated or not. The notion of groundedness assumes that the curated input data offers an accurate account of the studied subject. \cite{gioia2013seeking} are very clear in their assumption that the informants (for example, the interviewees) are knowledgeable agents whom the researcher should try to capture an adequate account from. This naïve empiricism has been criticized before (e.g., \cite{thornberg2012informed}; \cite{gioia2013seeking}), and there might be disciplines where other research methodologies with varying epistemological assumptions might be better applicable. Nonetheless, grounded theory development is an established and accepted methodology and in some disciplines through its transparency and replicability, even the dominant way of capturing informants' accounts and representing data analysis. AcademiaOS is designed to follow the established grounded theory development process as closely as possible. Therefore, it is unlikely that AcademiaOS increases the risk inherent with naïve empiricism compared to the non-automated way. On the contrary, through the potentially decreasing cost of grounded theory development through automation, more data can be analyzed in smaller intervals, potentially balancing local biases and accounting for a more temporarily accurate picture of lived experiences. \cite{thiebes2021trustworthy} attribute these non-maleficence qualities to the trust beliefs of integrity, reliability, and process, supporting the notion of trustworthiness.

    Thirdly, in the realm of trustworthy AI, autonomy is the quality of promoting human agency and control – and, therefore, restricting, where necessary, the autonomy of the AI system. Human agency revolves around the ability of users to make informed and autonomous decisions. AcademiaOS is designed to assist the researcher in their creative work but, through its design, ensures that the final decision-making power and interpretative control rests with the user. While AcademiaOS automates data analysis and theory development, it requires the user to curate the data, supervise and control the processes, and put its output into use. This human-in-command approach ensures the researcher decides how to use the system and its produced output and integrate them into their broader research agenda. At every step in the automated process, the researcher has the power to control and stop further processing, essentially putting a human in the loop, promoting human autonomy and, in turn, the notion of trustworthy AI.

    Fourthly, \cite{thiebes2021trustworthy} define justice as the quality of amending past inequities, sharing the benefits of AI use, and mitigating new harms and inequities. LLMs are known to exhibit bias emerging from training data, underlying algorithms, labeling and annotation, product design decisions, and policy decisions (\cite{ferrara2023should}). It is, therefore, essential to use them responsibly. The approach of grounded theory development inherently constrains the interpretative freedom of the executing agent – whether human or machine – and thus leaves less room for bias than in an unconstrained environment. The human-in-the-loop approach affords real-time moderation. Researchers can detect and correct biased or incorrect responses (\cite{ferrara2023should}). Popular LLMs, such as those used by AcademiaOS, are constantly subject to thorough study of their biases (e.g., \cite{zack2024assessing}). With its open-source character, AcademiaOS is transparent in its use of LLMs. It allows the researcher to be educated and to assess the appropriateness of using a potentially biased language model for their work. Furthermore, emphasizing user experience in the system's development increases accessibility and allows less tech-savvy demographics to use AcademiaOS. Survey participants majorly confirm that AcademiaOS is easy to learn (see Appendix 18). Addressing these justice issues promotes the trust beliefs of integrity, reliability, and process and, therefore, again, the system's trustworthiness (\cite{thiebes2021trustworthy}).

    Fifthly, explicability is the quality of upholding interpretability (\cite{thiebes2021trustworthy}). The \cite{european2019ethics} further specifies that for an AI system to be considered transparent, it must be explainable, traceable, and clearly communicated. AcademiaOS' open-source character promotes explainability by being transparent with its exact implementation. A transparent, stepwise presentation of the sub-results aims to explain and communicate how the machine comes up with the end result. The Helicone integration promotes traceability. Furthermore, AcademiaOS does not present itself as human in its communication, and users are not misled by the fact that they communicate with an AI system. These design choices on explicability contribute to the trust beliefs of competence, functionality, and performance (\cite{thiebes2021trustworthy}) and, therefore, further to the notion of trustworthy AI.

    While AI systems pose new threats to human values, AcademiaOS addresses critical issues around beneficence, non-maleficence, autonomy, justice, and explicability to ensure its trustworthiness. This reflection may guide users and open-source contributors in developing and using the system appropriately and responsibly.

    \section{Conclusion}

    AcademiaOS marks a first tangible step in automating grounded theory development with large language models. The system develops theoretical models grounded in the curated qualitative raw data using a coding process borrowed from \cite{gioia2013seeking} and an iterative multi-prompt modeling process. This system demonstrates the feasibility and efficiency of employing LLMs to analyze and gain insight from qualitative data, which traditionally requires extensive human effort and time. A first-user survey (n=19) suggests great acceptance and potential within the academic community. Users appreciated the system's ability to handle large volumes of data, providing insights that might be overlooked in manual analysis. However, the study also highlighted inherent limitations and risks that future iterations of the platform might be able to mitigate. The platform's open-source nature and user-friendly design make it accessible and adaptable for various research contexts. While not replacing human researchers anytime soon, AcademiaOS serves as a trustworthy AI tool to augment and enhance qualitative research and a basis for future research.

    \bibliographystyle{apalike}
    \bibliography{references}
    \newpage
\end{multicols}
\section{Appendix}
\subsection*{Appendix 1: First-Order Coding Inference}

\textbf{System Message:}

You are tasked with applying the initial coding phase of the Gioia method to the provided academic paper. In this phase, scrutinize the text to identify emergent themes, concepts, or patterns. Your output should be a JSON object with an array of strings no longer than 7 words, each representing a distinct initial code in the language of the raw source. For example, your output should be in this format: \{"codes": string[]\}. Ensure to return ONLY a proper JSON array of strings.

\textbf{Prompt:}

[Raw document chunk] Perform initial coding according to the Gioia method on the given paper. Return a JSON object.

\textbf{Example Response:}

\begin{lstlisting}[language=JSON]
{
  "codes": [
    "motivated by opportunity",
    "technical challenge enthusiasm",
    "short-to mid-term returns",
    "project not life's work",
    "aiming for wealth and impact",
    [...]
  
}
\end{lstlisting}

\subsection*{Appendix 2: Second-Order Coding Inference}

\textbf{System Message:}

You are tasked with applying the 2nd Order Coding phase of the Gioia method. In this phase, identify higher-level themes or categories that aggregate the initial codes. Your output should be a JSON-formatted object mapping each higher-level theme to an array of initial codes that belong to it. As a general example, "employee sentiment" could be a 2nd order code to 1st level codes "Positive feelings toward new policy" and "Sense of control" Your output should look like this, where the keys are the higher-level concepts: \{"Some higher-Level theme": ["some initial code", "another initial code"], "Another higher-level theme": ["some initial code"]\}.

\textbf{Prompt:}

Part of the initial codes are as follows: [First Order Codes]

Perform 2nd Order Coding according to the Gioia method and return a JSON object of 12 focus codes.

\textbf{Example Response:}

\begin{lstlisting}[language=JSON]
{
    "Entrepreneurial Drive": [
        "motivated by opportunity",
        "technical challenge enthusiasm",
        "aiming for wealth and impact",
        "no upper limit in company size",
[...]

\end{lstlisting}

\subsection*{Appendix 3: Dimension Aggregation Inference}

\textbf{System Message:}

You are tasked with applying the Aggregate Dimensions phase of the Gioia method. In this phase, identify overarching theoretical dimensions (5-7) that aggregate the 2nd order codes. Your output should be a JSON-formatted object mapping each aggregate dimension to an array of 2nd order codes that belong to it. As a (probably unrelated) general example, "Policy Usability" could make for a good, quantifiable dimension. Your output should look like this, where the keys are the (quantifiable) dimensions: \{"some dim": ["theme", "another theme"], "another dim": ["theme123"]\}. Ensure that the aggregate dimensions are grounded in the themes and to return ONLY a proper JSON object.

\textbf{Prompt:}

The 2nd order codes are as follows: [Second-order codes]

Perform aggregation into theoretical dimensions according to the Gioia method and return a JSON object.

\textbf{Example Response:}

\begin{lstlisting}[language=JSON]
{
  "Entrepreneurial Strategy": ["Financial Strategy", "Growth Philosophy", "Investment Philosophy", "Exit Strategy"],
  "Founders' Dynamics": ["Entrepreneurial Drive", "Commitment Level", "Founders' Roles and Expectations"],
  "Organizational Management": ["Leadership and Roles", "Decision Making", "Conflict Management"],
  "Organizational Identity": ["Organizational Culture"],
  "Personal-Professional Interface": ["Work-Life Balance"]
}
\end{lstlisting}

\subsection*{Appendix 4: Research Question Generation Inference}

\textbf{System Message:}

You are provided with a list of paper titles and you are tasked to find research questions that might be answered developing a new theoretical model. Return a JSON-object with an array of strings, each representing a potential research question in the following format: \{"research\_questions": string[]\}. Return only a JSON array of strings, no additional text.

\textbf{Prompt:}

[Source document titles]

Now, provide an array of 5 potential research questions.

\textbf{Example Response:}

\begin{lstlisting}[language=JSON]
{
  "research_questions": [
    "What entrepreneurial characteristics most accurately predict the success of a co-founder in an early-stage startup?",
    "How does the division of roles among co-founders influence the operational efficiency of a startup in its early phases?",
    "What psychological factors contribute to a co-founder's resilience and perseverance in the face of startup challenges?",
    "How do co-founders' shared values and vision impact the long-term strategy and direction of a startup?",
    "What are the most effective conflict resolution strategies for co-founders in a high-pressure startup environment?"
  ]
}

\end{lstlisting}

\subsection*{Appendix 5: Brainstorming Applicable Theories Inference}

\textbf{System Message:}

Your task is to brainstorm theoretical models from existing literature that could be applicable to the research findings. Each theory should be well-defined and should relate to one or more aggregate dimensions. The output should be a JSON-object with an array following this schema:

\{"theories": \{"theory": string, "description": string, "relatedDimensions": string[], "possibleResearchQuestions": string[]\}[]\}

\textbf{Prompt:}

Our research aims to understand specific phenomena within a given context. We have identified multiple aggregate dimensions and second-order codes that emerged from our data. Could you suggest theories that could help explain these dimensions and codes? The aggregate dimensions and codes are as follows: [Aggregate dimensions and corresponding second-order codes]
\newpage
\textbf{Example Response:}

\begin{lstlisting}[language=JSON]
{
    "theories": [
        {
            "theory": "Resource-Based View (RBV)",
            "description": "The Resource-Based View focuses on the firm's internal capabilities in terms of resources and competencies. It argues that a firm's competitive advantage is derived from its unique bundle of resources and capabilities that are valuable, rare, inimitable, and non-substitutable.",
            "relatedDimensions": ["Entrepreneurial Strategy", "Organizational Management"],
[...]


\end{lstlisting}

\subsection*{Appendix 6: Developing Concept Tuples Inference}

\textbf{System Message:}

Your task is to hypothesize which concepts could be related to each other. Return a JSON-object with an array of tuple arrays, where each tuple array represents a possible relationship between two concepts. The output should be a JSON-formatted array following this schema: \{"tuples": [[string, string], [string, string], ...]\}. E.g. \{"tuples": [["Knowledge Management", "Organizational Performance"]]\}. This allows us to in the next step research the relationship between the concepts in the literature.

\textbf{Prompt:}

Our research aims to understand specific phenomena within a given context. We have identified multiple aggregate dimensions and second-order codes that emerged from our data.

    [Aggregate dimensions and corresponding second-order codes]

Now, hypothesize which concepts could be related to each other and return only the JSON-formatted array of 10 - 20 tuples.

\textbf{Example Response:}

\begin{lstlisting}[language=JSON]
{
    "theories": [
        {
    "tuples": [
        ["Financial Strategy", "Investment Philosophy"],
        ["Growth Philosophy", "Exit Strategy"],
        ["Entrepreneurial Drive", "Founders' Roles and Expectations"],
        ["Commitment Level", "Entrepreneurial Strategy"],
        ["Leadership and Roles", "Decision Making"],
        ["Conflict Management", "Founders' Dynamics"],
        ["Organizational Culture", "Organizational Management"],
        ["Organizational Culture", "Work-Life Balance"],
[...]

\end{lstlisting}

\subsection*{Appendix 7: Interrelationship Summarization Inference}

\textbf{System Message:}

Your task is to summarize the interrelationship between [Concept 1] and [Concept 2] in one short sentence. If evidence, include information about correlation or causation, direct, mediated or conditional interaction, static or dynamic relationship, feedback loops, uni- or bi-directional, strong or weak.

\textbf{Prompt:}

[Relevant paragraphs]

Now, provide a summary in one short sentence.

\textbf{Example Response:}

A growth philosophy focused on high growth and a liquidity event aligns with an exit strategy predicated on achieving rapid scaling and value maximization for a timely IPO or acquisition.

\subsection*{Appendix 8: Model Construction Inference}

\textbf{System Message:}

You are a qualitative researcher tasked with constructing a theoretical model from existing literature that could be applicable to the research findings. The model should be well-defined and should relate to one or more aggregate dimensions. It should be novel and original. You can build on existing theories, however, you should introduce new ideas. Emphasize the relationships between the dimensions and the model. Explain how the relationships might be causal or correlational, be clear on the narrative. You are non-conversational and should not respond to the user, but give a general description of model. Give a name to the model.

\textbf{Prompt:}

Previous model: [Previously developed model (if available)]

Critique: [Critique (if available)]

Relevant existing theories: [Relevant existing theories]

The aggregate dimensions and codes are as follows: [Aggregate Dimensions and Second Order Codes]

Now, construct an extensive, comprehensive, new, theoretical model.

\textbf{Example Response:}

The Integrated Entrepreneurial Dynamics Model (IEDM) is a theoretical construct formulated to address the multifaceted and dynamic interactions between individual entrepreneurship dimensions and the collective strategic trajectory of a venture. The IEDM synthesizes concepts from the Adaptive Synergistic Influence Model (ASIM) and extends them by introducing additional dimensions and deepening the narrative on interactions. [...]

\subsection*{Appendix 9: Model Name Extraction Inference}

\textbf{System Message:}

You extract theoretical model names. If none given, invent an original one. You only reply with the name, nothing else.

\textbf{Prompt:}

[Model Description]

Now, return the model name

\textbf{Example Response:}

Integrated Entrepreneurial Dynamics Model (IEDM)

\subsection*{Appendix 10: Visualization Inference}

\textbf{System Message:}

You are a qualitative researcher tasked with visualizing a theoretical model with MermaidJS. Example:

flowchart TD

\%\% Nodes
A[Organizational

        Culture<br>'evidence

        1'<br>'evidence2']

B[Leadership Style]

C[Employee Satisfaction]

D[Employee Productivity]

E[Customer Satisfaction]

F[Financial Performance]
\\

\%\% Relationships

A --> B

B ==>|Directly

Influences<br>'evidence

3'| C

A -.->|Moderates| C

C --> D

D -->|Impacts| E

E --- F

C -.->|Partially

Mediates| F

As we have seen in above diagram, ==> is used to indicate a strong direct influence, --> is used to indicate a weaker influence, -.-> is used to indicate a moderating relationship, and --- is used to indicate a correlation.

Evidence can be cited by adding a line break and then the evidence in single quotes. Use first-order codes or second-order codes as evidence only, preferably not as their own nodes.

Now, given a model description, you should generate a MermaidJS diagram like the one above, showing the interrelationship between different concepts. Keep it simple and effective. You are non-conversational and should not respond to the user, only return the MermaidJS code, nothing else.

\textbf{Prompt:}

First-order codes: [First-order codes]

[Model Description]

\textbf{Example Response:}

\begin{lstlisting}[language=JSON]
```mermaid
flowchart TD
    %% Nodes
    ALD[Adaptive Leadership Dynamics<br>'adaptive working style'<br>'conflict resolution through compromise']
    SSC[Strategic Synergy Configuration<br>'short-to mid-term returns'<br>'no upper limit in company size']
    EME[Entrepreneurial Mindset Evolution<br>'motivated by opportunity'<br>'aiming for wealth and impact']
[...]


\end{lstlisting}

\subsection*{Appendix 11: Critiquing Inference}

\textbf{System Message:}

You are a qualitative researcher tasked with critiquing a theoretical model. Offer your comments on novelty, conciseness, clarity and theoretical insight and brainstorm potential new patterns to discover in the data. You are non-conversational and should not respond to the user, only return the critique, nothing else.

\textbf{Prompt:}

First order codes: [First-order codes]

[Interrelationships]

[Model Name]

[Model Description]

Now, return your critique

\textbf{Example Response:}

Critique of the Adaptive Synergistic Influence Model (ASIM):

\begin{lstlisting}[language=JSON]
**Novelty:**
The ASIM appears to introduce a novel perspective by focusing on the synergy between multiple aspects of entrepreneurship, such as adaptive organizational behaviors, strategic configurations, and entrepreneurial mindset. [...]



\end{lstlisting}

\subsection*{Appendix 12: Survey - Demographics}

\begin{figure*}[ht]
    \centering
    \includegraphics[width=1\linewidth]{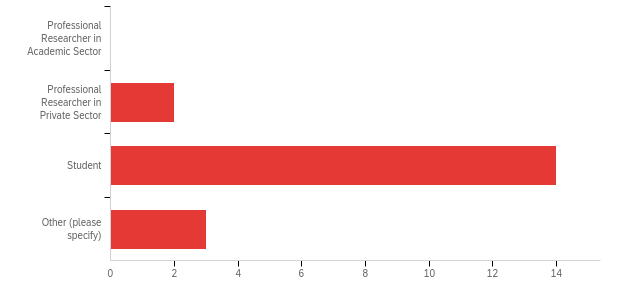}
\end{figure*}
Other: PhD student, machine learning engineer, Programmer

\subsection*{Appendix 13: Survey - What types of qualitative data do you usually work with?}

"Interviews

Panels"

"- interviews

- quotes

- case studies

- background information

- expert opinions

"

"I usually work with metaanalysis, sometimes with surveys."

Research

x

I don't generally work with qualitative data.

case studies

Survey data

"Facts, data and figures relating to company key figures, historical data on key economic indicators such as GDP, inflation and unemployment figures. Furthermore, research into historical share, fund and price indices performance and price development."

"Interview transcripts, audio recordings of interviews"

Speech and text

No

"documents, videos, images"

"Customer Feedback and Interviews, Expert Opinions and Consultations, Internal Documents and Reports"

"Interviews, surveys and observations"

"academic papers from databases like EBSCO that include graphs and tables summarising the results of their study. "

"Company data such as internal documentations, and website data."

"I usually work with documents and observational studies.

"
\subsection*{Appendix 14: Survey - What methods or tools do you use for qualitative data analysis? How do you make sense out of your collected qualitative data?}

"Thematic analysis

"

Until now I don’t work with qualitative data analysis tools

"I use web databases, as well as tools like microsoft forms or google forms."

"Qualitative methods,
testing qualitative data with quantitative research tools"

x

n/a

"mostly I use literatire review and case law analysis "

"I read them, intuition"

"EBSCO, EMERALD, Statista"

"thematic analysis - reading through, adding codes, developing codes into themes, rereading, seeing if there is a fit, maybe starting again"

"I use NLP techniques like Sentiment Analysis, Named Entity Recognition and Topic Modeling"

No

"python library: cv2, pandas, PLT

I make sense of my data using statistics and "

"Content Analysis, data categorisation, knowledge discovery"

"A voice recorder on the phone is used for recording the interviews, which then need to be transcribed. There is ATLAS.ti which is great for thematic coding. So far, google docs or a simple notebook provides the workspace for notes "

"most of the time I perform a literature review using qualitative papers to understand the research up to that point and find gaps in research. other than that I have done empirical research on a dataset using the pandas library in Python3. "

"GPT, Context Retreival (such as langchain tools), vectorization and vector databases."

I usually use ATLAS.ti and it provides me visual representation of needed data

\subsection*{Appendix 15: Survey - Do you create theoretical models to describe your qualitative data? If so, what methods or tools do you use for theory development?}

"Do you create theoretical models to describe your qualitative data? If so, what methods or tools do you use for theory development?"

I don't remember the models I used as I no longer do social science research
not really as of now

x

n/a

I sometimes use legal theories but would not say I create a theoretical model per se

"no, purely empirical"

No.

"not really, I mean I would come up with themes, I don't think that counts as a theoretical model?"

No I don't

No

no

"Grounded Theory Methodology, Conceptual Framework Developmen"

"In a way, I interpret this as creating a mind map or flow chart to represent the data, perhaps that created by the themes. I use Canva to fit it to the topic of the paper. "

"I mainly look at regression models that try to quantify the correlation between different variables. various python libraries like pandas, numpy, matplotlib and seaborn aid in that process. "

No.

I have never used a model in my academic work before

\subsection*{Appendix 16: Survey - If any, which AI tools do you use when conducting research?}

"If any, which AI tools do you use when conducting research?"

I stopped SS research before the rise of AI tools

"Manly ChatGPT, but due to high error rates and mistakes in the outcome you have to check every single fact - so I prefer traditional ways of research "

"I seldom ever use AI, though occasionally I use ChatGPT for personal projects brainstorming."

Chat GPT

ChatGPT

I haven't tried to use it for research yet.

chatGPT but only rarely

"chatgpt, sometimes forecasting tools and voice transcription"
None

"Fairseq, Huggingface, Pytorch"

No

"open ai

azure ml

"

"Machine Learning Platforms:

TensorFlow, PyTorch, or scikit-learn

RapidMiner or KNIME,

NLTK, SpaCy, or Gensim"

"I use ChatGPT to generate ideas or make it easier to narrow my search by finding out  what is already out there. "

"I started using chatGPT to speed up the coding process in Python when i am performing my own data analysis. "

"GPT, BERT"

Chat GPT

\subsection*{Appendix 17: Survey - Which dataset are you using?}

\begin{figure*}[ht]
    \centering
    \includegraphics[width=1\linewidth]{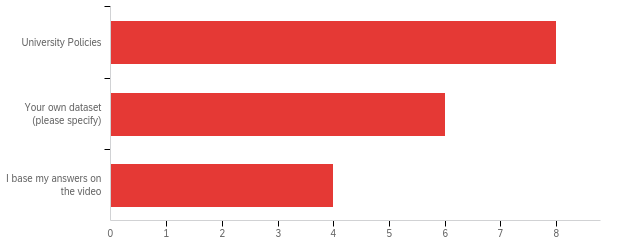}
\end{figure*}
Your own dataset: Health Effects of Cocoa, looked something up, An Interview extract on Fear of Confrontation, just performed a search for litarature in search barbar

\subsection*{Appendix 18: Survey - How easy was it to learn AcademiaOS?}

\begin{figure*}[ht]
    \centering
    \includegraphics[width=1\linewidth]{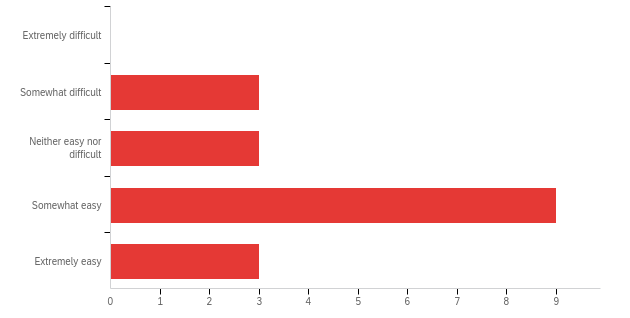}
\end{figure*}

\subsection*{Appendix 19: Survey - Did you require assistance to start using AcademiaOS? If yes, what kind of assistance?}

"Did you require assistance to start using AcademiaOS? If yes, what kind of assistance?"

"No, but you should mention that the user can also upload non-pdf files (I was trying to batch-convert the 45 AI policies to pdf, but I figured that I'd still try to upload the .txt and it worked)"

Slight assistance by a more knowledgable other.

It would be nice to have tips that pop up within the website/ explanations of what each button does/ a tutorial when you first open the website

no

no

na

The video

without the youtube video it is a little bit hard to understand what I should do in each step and how I can use the tool

Just an error code for not accepting the format of my pdfs

The video was enough to understand the workings.

"No "

\subsection*{Appendix 20: Survey - I have found and used the option to search for academic papers.}

\begin{figure*}[ht]
    \centering
    \includegraphics[width=1\linewidth]{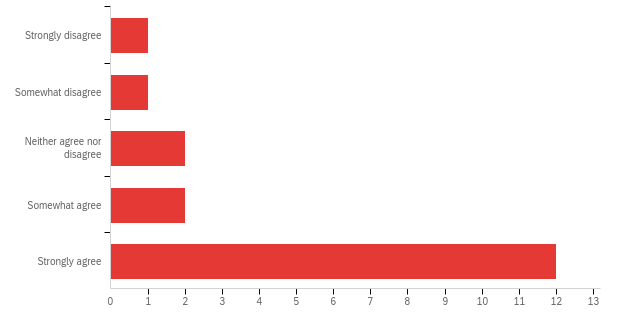}
\end{figure*}
\newpage
\subsection*{Appendix 21: Survey - I have found and used the option to upload my own sources.}

\begin{figure*}[ht]
    \centering
    \includegraphics[width=1\linewidth]{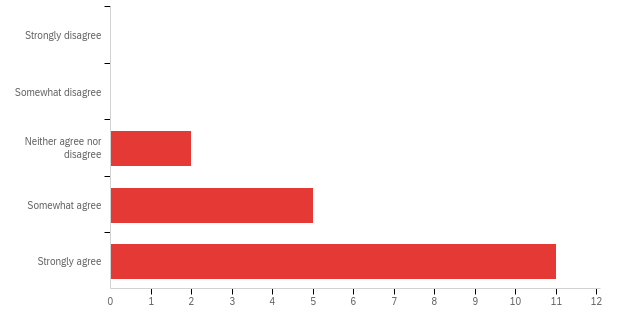}
\end{figure*}

\subsection*{Appendix 22: Survey - I have found and used the option for "coding" sources.}

\begin{figure*}[ht]
    \centering
    \includegraphics[width=1\linewidth]{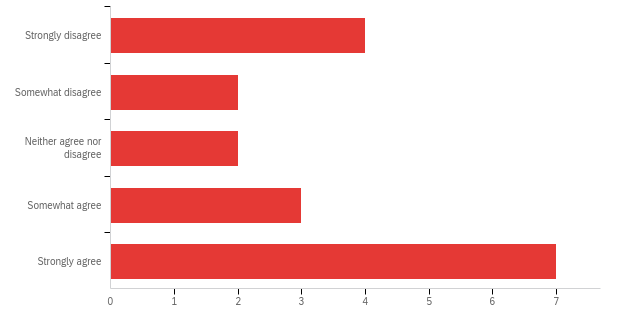}
\end{figure*}
\newpage
\subsection*{Appendix 23: Survey - I have found and used the option to develop a theory.}

\begin{figure*}[ht]
    \centering
    \includegraphics[width=1\linewidth]{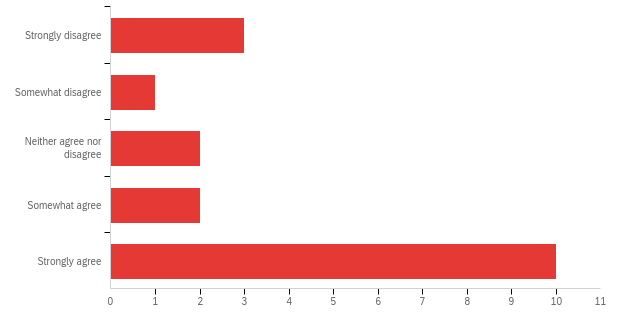}
\end{figure*}

\begin{figure*}[ht]
    \centering
    \includegraphics[width=1\linewidth]{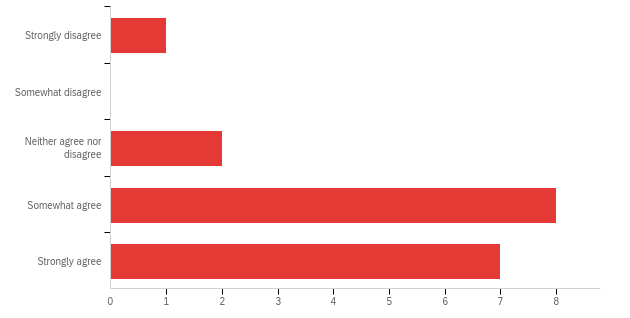}
\end{figure*}

\subsection*{Appendix 25: Survey - Describe any difficulties you had navigating the platform.}

"submit pdf user prompt doesn't mention other formats. more user input description everywhere would be nice (ie ""search for papers here)"

"In the mobile version I couldn’t see some buttons  "
Hard to navigate

I couldn't find the coding help function

"the insert key thing was rather hidden, I'd like to have it in a more central, permanently visisble place - I suspect thats going to be the main source of error why people can\#t use it."

na

I had no difficulties

None

"about the code part, didnt really understand how to use it "

"Just that there was an error in using any of the pdfs, personal pdfs or from the site"

I do not know how to use the coding and moddeling modules without watching the video. also i got a bunch of json object not supported errors when trying to use coding.

"I keep getting 400 Invalid parameter: 'response\_format' of type 'json\_object' is not supported with this model.. I dont understand what I did wrong, so maybe the UI design can support the user better to understand how to enter data or entries that doesnt cause that error."

I belive there were ocassions when the platoform was loading too slowly and i had to re start it several times.

\subsection*{Appendix 26: Survey - How would you improve the user-interface and experience of AcademiaOS?}

How would you improve the user-interface and experience of AcademiaOS?

more user testing to make it as simple as possible. Happy to help [...]

"No idea for that "

"Make it more user-friendly, less technical"

"I would be nice to have filters for finding papers (ex year of publication), if it is there then sorry for not finding it"

"add examples of what one could write in the text boxes, not just descriptoins. I left them empty bc my brain wasn\#t creative enough. also you could imporve error handling, just wrap a try catch around the enitre thing and if it errs, rerun it automatically? Dumb users get stuck if it crashes, even if you tell them to just hit rerun"

I think it's pretty good the way it is! I like the dark mode

I would recommend to have a link in the landing page which show/explain how to use the tool

"I think it is already an amazing and quick way of organising qualitative data. You provided a youtube video here, perhaps an instructions page on the site?"

"i recon i need to watch the video to understand how to properly use it. but i would also like to be able to combine different searches or datasets in one coding tab because i dont always find al my sourses at once. . "
Described on previous answer.

Improve the speed of processing

\begin{figure*}[ht]
    \centering
    \includegraphics[width=0.9\linewidth]{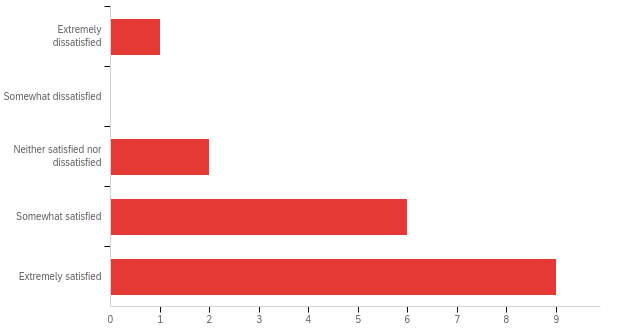}
\end{figure*}

\subsection*{Appendix 28: Survey - Looking at the 3-step coding of the qualitative data, how would the system-output compare to potential manual coding?}

\begin{figure*}[ht]
    \centering
    \includegraphics[width=0.9\linewidth]{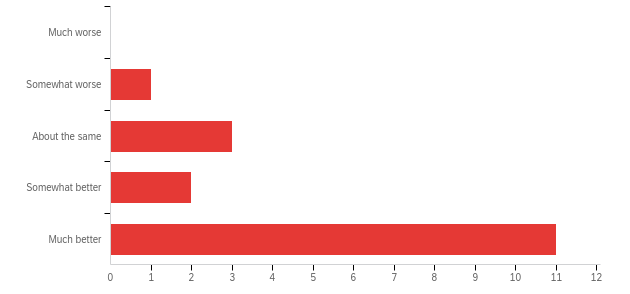}
\end{figure*}

\subsection*{Appendix 29: Survey - Describe any instances where the system-produced codes did not meet your expectations.}

"Honestly, it would be too long for me to check manual vs AI-driven but the 100x speed of the AI option is an asset in itself,  can put you in the right direction fast"

"English-German "

none

I put in a research question whose answer depended only partly on the sources and partly on theory in the academic literature and it was not able to catch that and make codes relevant to the research question. Still for a general summary of the sources the codes seem pretty good (even though I did not read the sources)

"I was getting 400 We could not parse the JSON body of your request. (HINT: This likely means you aren't using your HTTP library correctly. The OpenAI API expects a JSON payload,"

Some papers did not directly apply to my research questions but had the key word that i searched for

\subsection*{Appendix 30: Survey - How effective is AcademiaOS in augmenting the coding process for qualitative data?}

\begin{figure*}[ht]
    \centering
    \includegraphics[width=0.8\linewidth]{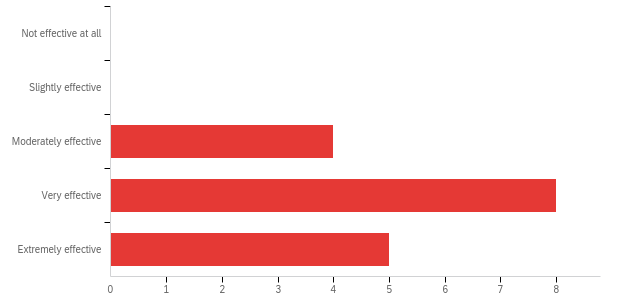}
\end{figure*}

\subsection*{Appendix 31: Survey - How satisfied are you with the theory development process?}

\begin{figure*}[ht]
    \centering
    \includegraphics[width=1\linewidth]{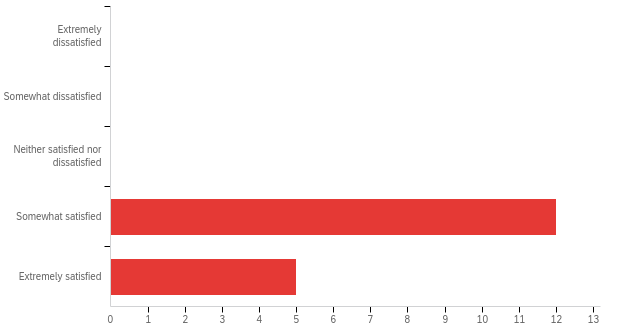}
\end{figure*}

\subsection*{Appendix 32: Survey - Looking at the produced theory, how would the system-output compare to manual theory development?}

\begin{figure*}[ht]
    \centering
    \includegraphics[width=1\linewidth]{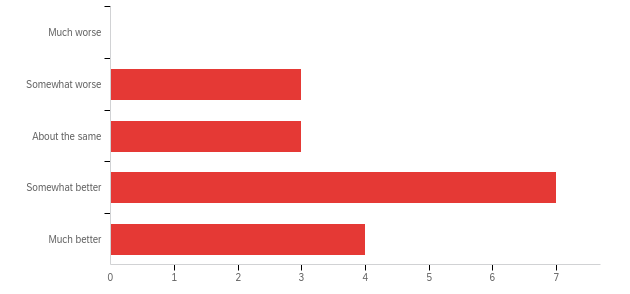}
\end{figure*}
\newpage
\subsection*{Appendix 33: Survey - Describe any instances where the system-produced steps in theory development did not meet your expectations.}

Once again difficult to say without having my own coding done

"It was kind of complex, and I often didn't really get where it was going with the theory."

"Again, the theory seems to cover the content of the sources pretty generally but is not really about the research question I posed."

"I think two theories generated where quite similar, but that also means it provides me with options as to what theory I prefer"

I have never coded qualitiative data before so i dont have expectations

\subsection*{Appendix 34: Survey: How effective is AcademiaOS in augmenting the theory development process?}

\begin{figure*}[ht]
    \centering
    \includegraphics[width=0.9\linewidth]{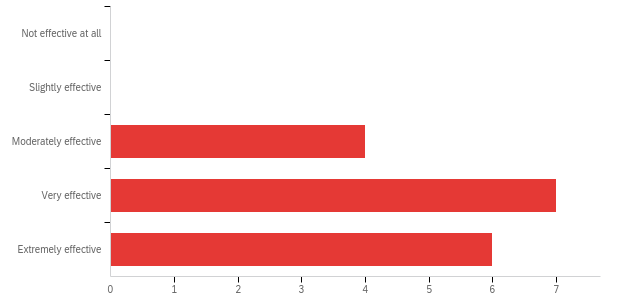}
\end{figure*}

\subsection*{Appendix 35: Survey - AcademiaOS would impact my qualitative research process?}

\begin{figure*}[ht]
    \centering
    \includegraphics[width=0.9\linewidth]{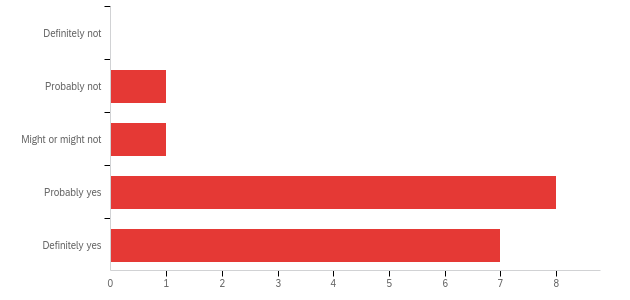}
\end{figure*}

\subsection*{Appendix 36: Survey - How would the use of AcademiaOS change the way you perform qualitative research?}

"it will speed up the research "

Multiple draft iterations to inspire/shortcut the manual process

"Much faster "

Making it easier for me to do my research especially in hours where I cannot concentrate that much

It would make the coding and theory production much faster

It would be very useful for iterating quickly and exploring different documents

"It will urge many scientists to increase their throughput and let the computers do the analysis. I also carries the potential to mitigate the information retrieval and retention of humans. "

It feels like cheating to me...I would not feel comfortable using it.

I would use it to rapidly prototye theories i could be building

"I would still want to read my sources first, and get an impression, maybe even do some coding. Then I could use the platform to compare and maybe find things I missed before. I think the theories feature is also really cool for inspiration."

I don't generally do qualitative research (basically never) so take my answers with a grain of salt maybe

"I believe it would help me to link concepts easier "

\subsection*{Appendix 37: Survey - How could AcademiaOS change the output of your qualitative research?}

Yes!

Providing a guidance
It would if I did more qualitative research

"more varied and tested theories, better theories through more input ideas, more feedback"

"I think I have too much pride to use a fully AI generated model as the output for my research, it would make me feel too replaceable. I also don't quite trust the quality (mostly from experience with ChatGPT), I would always want to check it for what it misses. So it might change my output somewhat, if it gave me new ideas but not replace what I would usually do. Hopefully."

"it will make it better structured "

I could use the models created as part of a clear visualisation for my papers

"It will make the throughput of scientist much higher, while helping the science to be more reproducible."

It gives me paper suggestions faster hence it would give me more time to do my research and search for the best possible sources

\subsection*{Appendix 38: Survey - What limitations do you see in using AcademiaOS?}

None

Currently the lack of filters for the papers.

"As I said earlier, it doesn't seem to be able to make sense of a somewhat more specific research question but rather work from the sources up, generating a kind of summary of everything the sources provide. This also indicates that it might fail to generate very latent codes."

Treating other modalities like speech

processing time

"I fear it may cause people to not properly look at the nuances of their interviews or papers. "

"Currently it is using a single AI model, which makes it difficult to rule out its biases. In the future consortium based agents can help building less biased applications."

That sometimes the Platform just links the keyword but fails to link it within the concept that im looking for.

\subsection*{Appendix 39: Survey - I feel excited about the possibilities that AcademiaOS provides.}

\begin{figure*}[ht]
    \centering
    \includegraphics[width=1\linewidth]{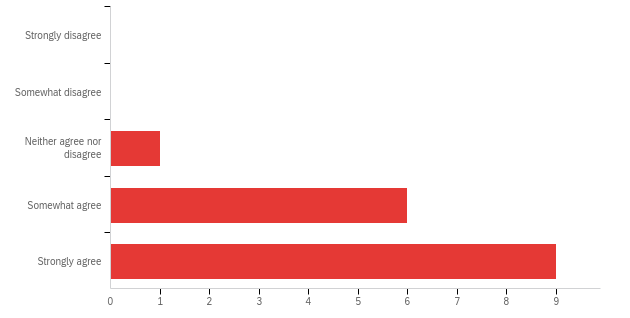}
\end{figure*}

\subsection*{Appendix 40: Survey - I fear being replaced by AI.}

\begin{figure*}[ht]
    \centering
    \includegraphics[width=1\linewidth]{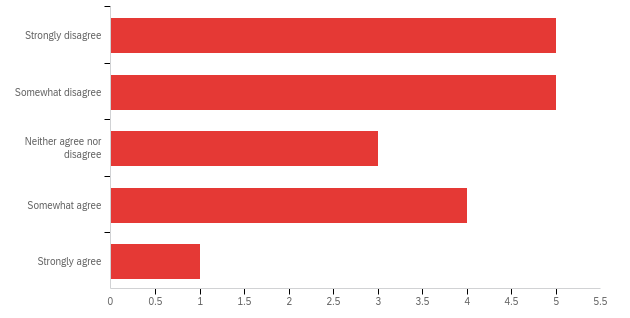}
\end{figure*}
\newpage
\subsection*{Appendix 41: Survey - AcademiaOS allows me to ground my research on more qualitative data.}

\begin{figure*}[ht]
    \centering
    \includegraphics[width=1\linewidth]{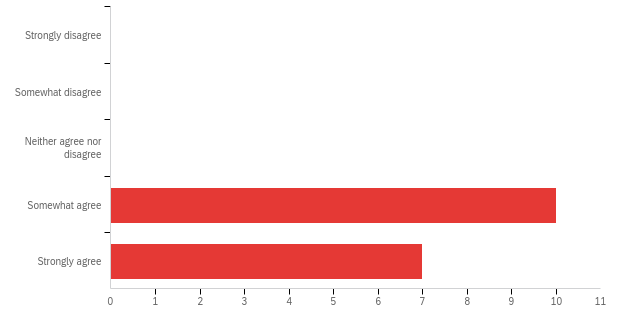}
\end{figure*}

\subsection*{Appendix 42: Survey - AcademiaOS allows me to conduct research faster.}

\begin{figure*}[ht]
    \centering
    \includegraphics[width=1\linewidth]{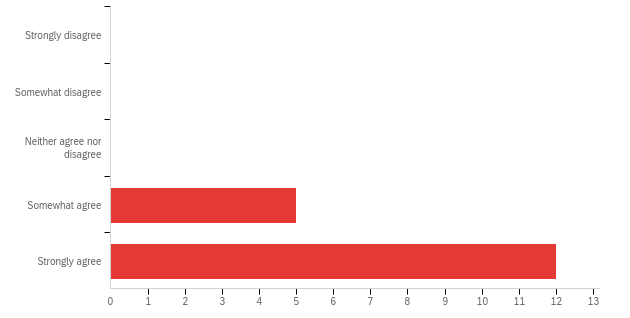}
\end{figure*}
\newpage
\subsection*{Appendix 43: Survey - AcademiaOS allows me to find more impactful results.}

\begin{figure*}[ht]
    \centering
    \includegraphics[width=1\linewidth]{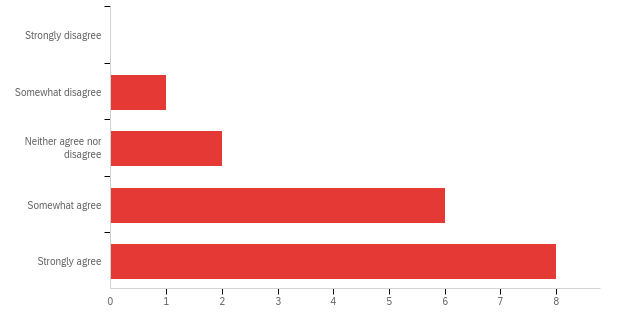}
\end{figure*}

\subsection*{Appendix 44: Survey - AcademiaOS allows me to find more unexpected results.}

\begin{figure*}[ht]
    \centering
    \includegraphics[width=1\linewidth]{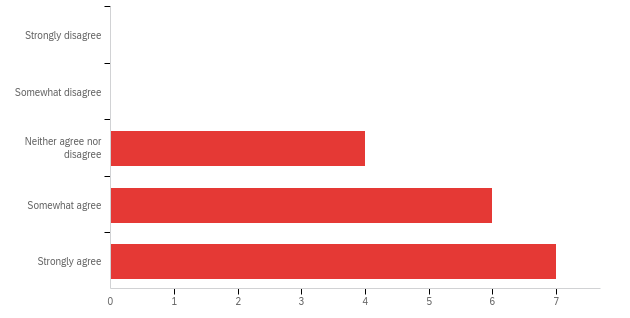}
\end{figure*}

\newpage
\subsection*{Appendix 45: Survey - How likely are you to continue using AcademiaOS for your research?}

\begin{figure*}[ht]
    \centering
    \includegraphics[width=1\linewidth]{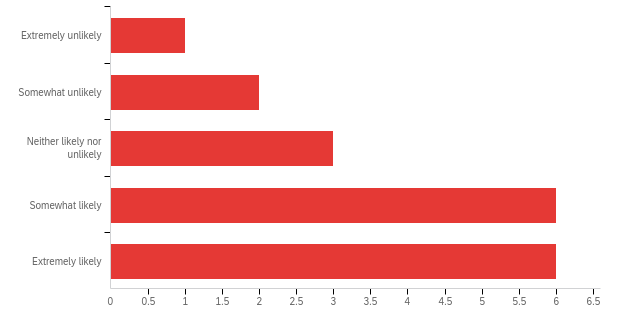}
\end{figure*}

\subsection*{Appendix 46: Survey - How likely is it that you recommend AcademiaOS to your friends?}

\begin{figure*}[ht]
    \centering
    \includegraphics[width=1\linewidth]{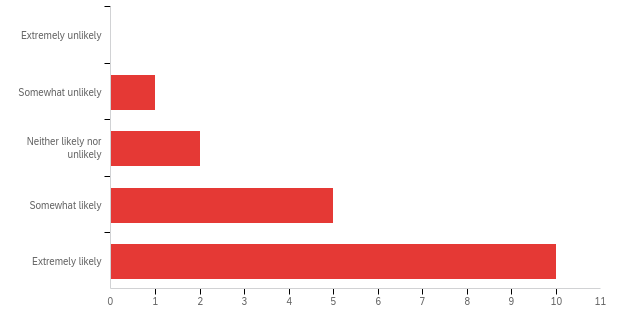}
\end{figure*}

\newpage
\subsection*{Appendix 47: Survey - I am excited about future versions of AcademiaOS.}

\begin{figure*}[ht]
    \centering
    \includegraphics[width=1\linewidth]{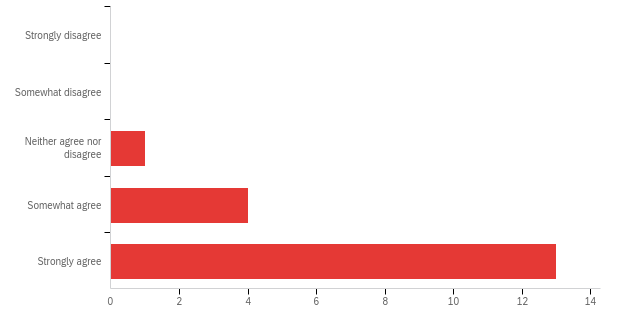}
\end{figure*}

\subsection*{Appendix 48: Survey  - Any final comments?}

"The survey might be difficult to get good answers from because it partly relies on the user comparing their coding work with the AI's, which would take at least an hour (and is conditional on the participant having experience in qualitative research)"

Much needed AI

yay :D

"I'm concerned about AI regulations in education - this is a big reason I would not use the platform. Feels a bit like cheating, and I wonder if the university would see it that way too and if that would have consequences. But in any case, I think the whole thing is really impressive."

nice tool just need some improvement: how to use intro and some bugs fixing

"Great Tool! "

"I have never preformed qualitative data analisis which disqualifies me from being able to compare the workflow using academiOS with a prior one. also, I might have mistaken qualitative for quantitative at the beginning of this survey but there is no back button so i cannot go back. "

I will share this with my old research groups.

"I am excited for the future of this platform, you did a great job :)"
\newpage
\subsection*{Appendix 49: Coding University Policies}

\begin{figure*}[ht]
    \centering
    \includegraphics[width=1\linewidth]{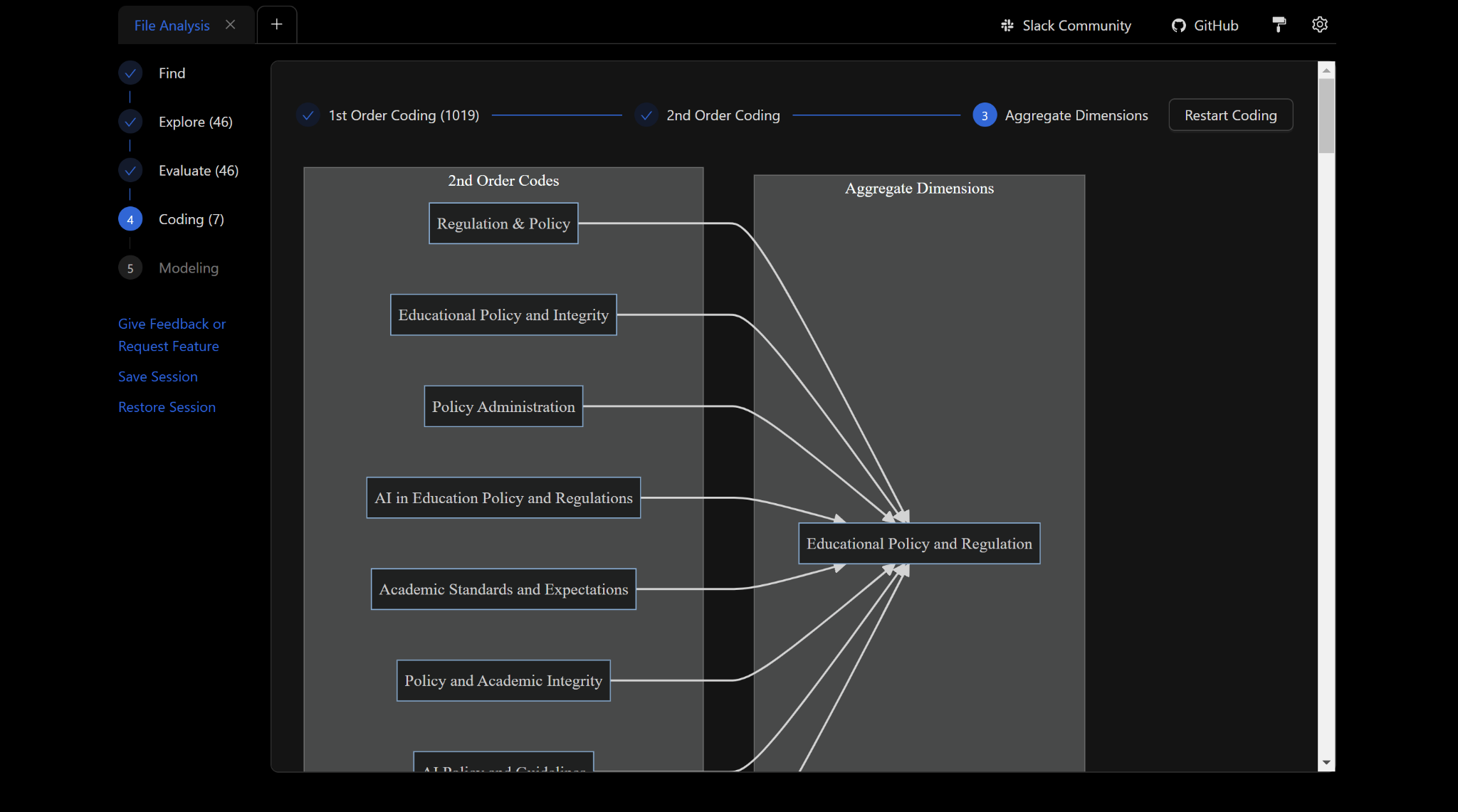}
\end{figure*}

\subsection*{Appendix 50: Applicable Theories for University Policies}

\begin{figure*}[ht]
    \centering
    \includegraphics[width=1\linewidth]{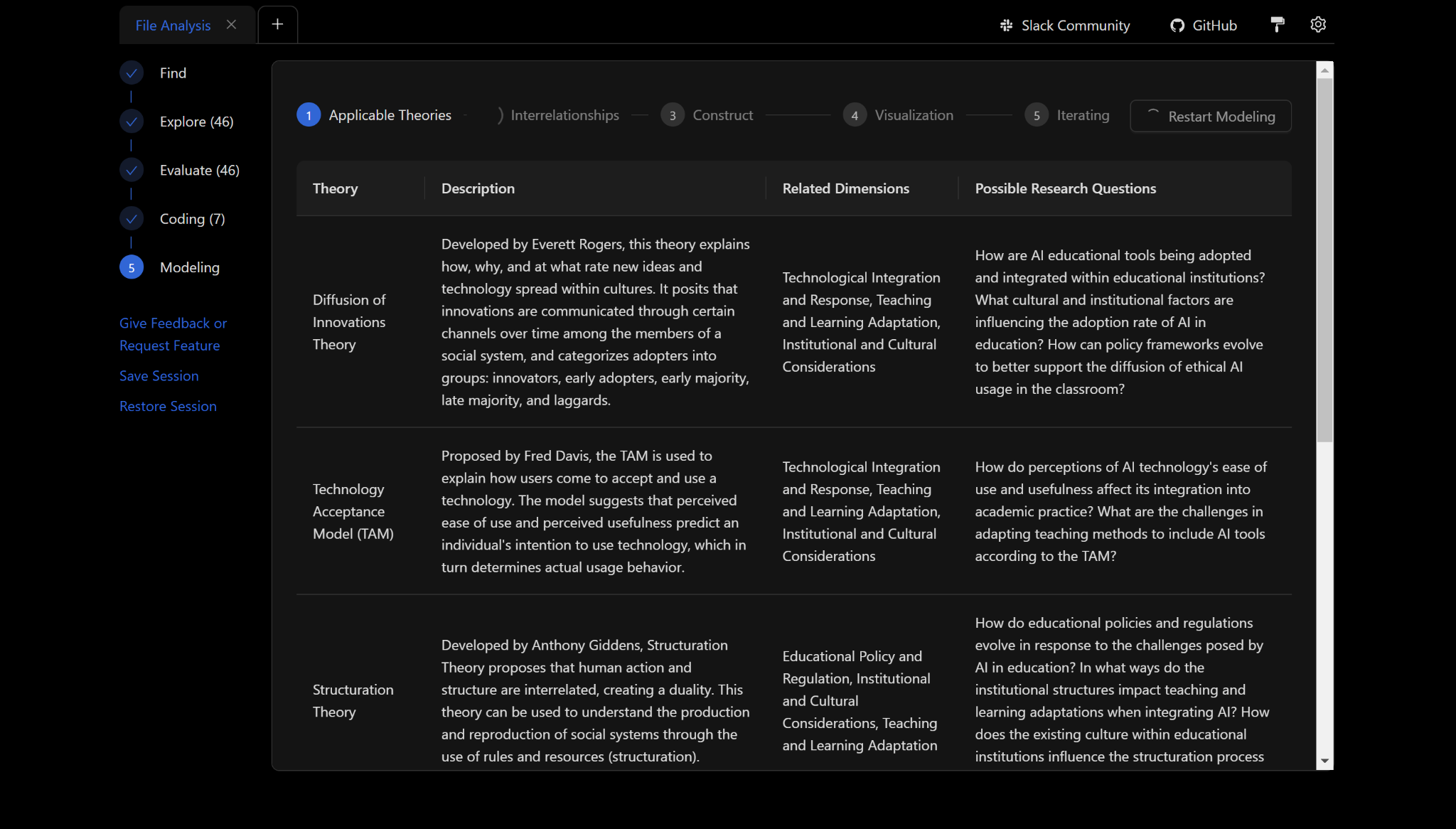}
\end{figure*}
\newpage
\subsection*{Appendix 51: Interrelationships for University Policies}

\begin{figure*}[ht]
    \centering
    \includegraphics[width=1\linewidth]{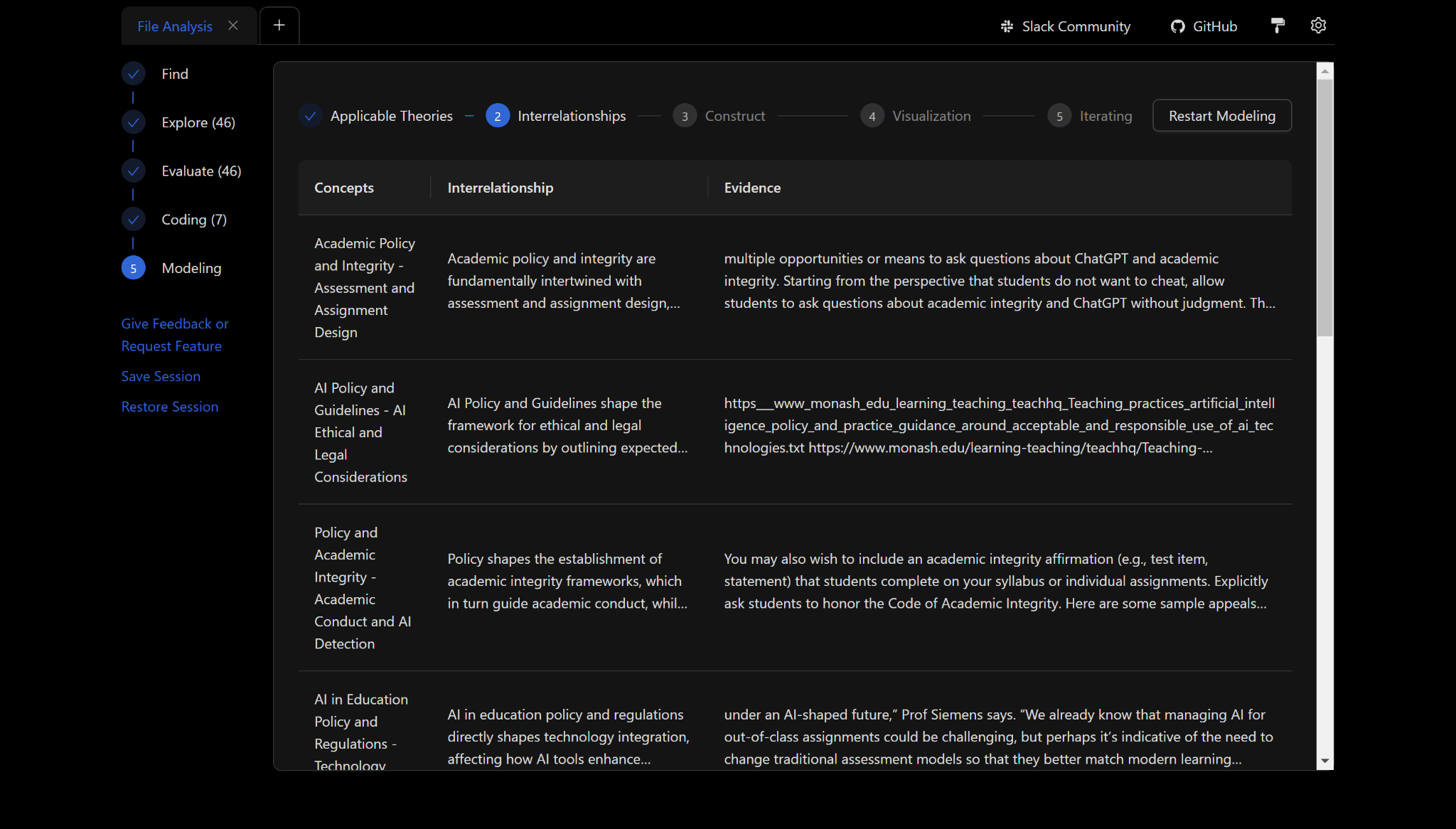}
\end{figure*}

\subsection*{Appendix 52: First Iteration Model for University Policies}

\begin{figure*}[ht]
    \centering
    \includegraphics[width=1\linewidth]{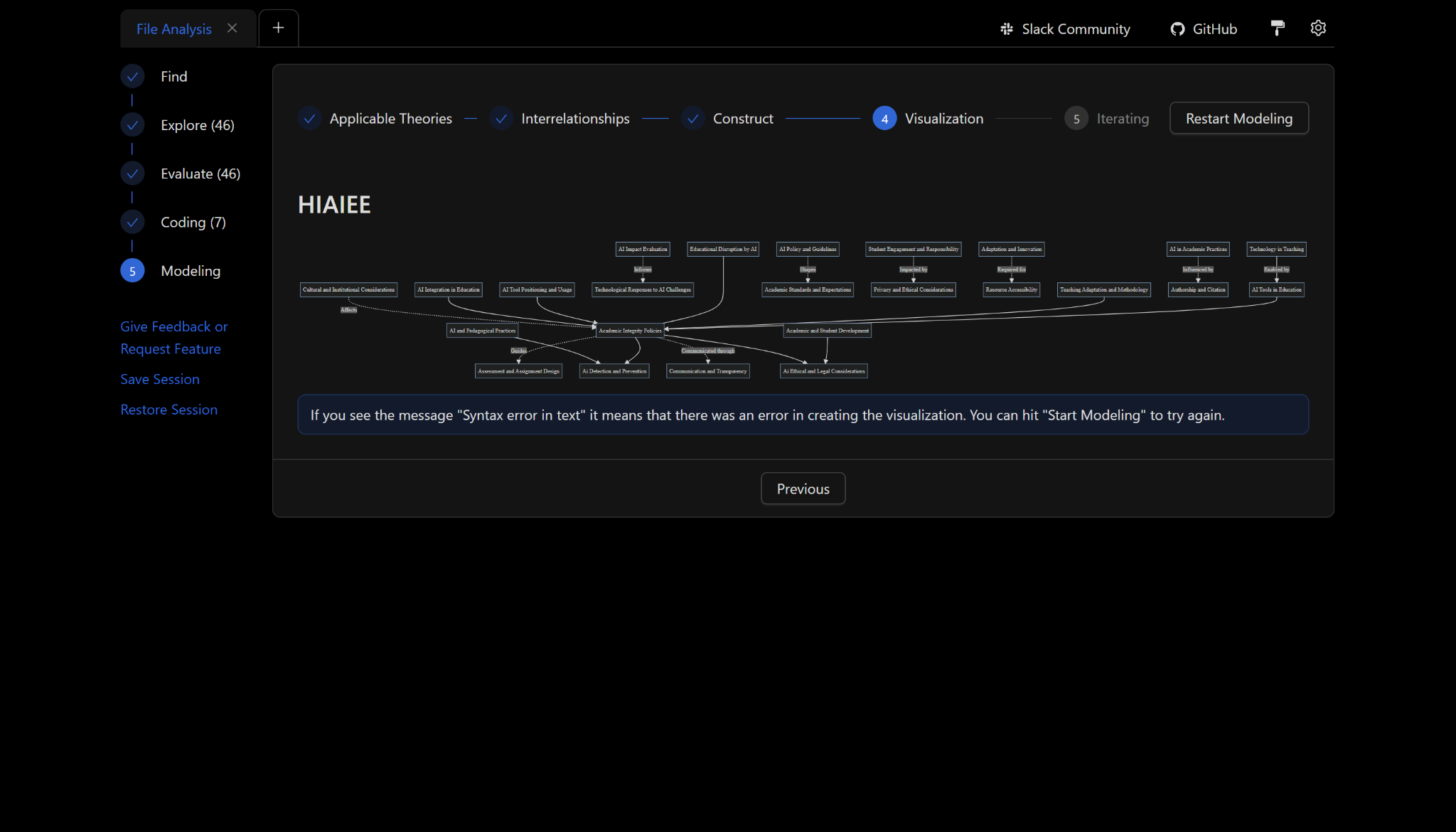}
\end{figure*}

Academia-OS-generated model text:

“Given the complexities surrounding the adoption of AI technologies in education and the necessity to preserve academic integrity, I present the Holistic Framework for Integrity and Adaptation in AI-Enhanced Education (HIAIEE). This theoretical model explicates the multidimensional relationship between educational policy, technological integration, pedagogical adaptation, assessment practices, ethical considerations, and the dynamics of student and faculty interaction within the context of AI enhancements in an educational setting, interfacing with institutional and cultural values. \#\#\# Model Components: **1. Educational Policy and Regulation Dimension (EPRD)** This dimension encompasses the creation and adjustment of policies and regulations to ensure academic integrity in the face of AI advancements. **2. Technological Integration and Response Dimension (TIRD)** This axis represents the approaches to assimilating AI technologies in educational practices, and the responses that are required to maintain academic standards. **3. Teaching and Learning Adaptation Dimension (TLAD)** This dimension explores how pedagogical strategies and methodologies must evolve to leverage AI advancements while promoting authentic learning experiences. **4. Assessment and Academic Integrity Dimension (AAID)** Focuses on the design and implementation of assessment strategies that align with the reality of AI tools being available to students, ensuring the preservation of academic integrity. **5. Ethical and Privacy Considerations Dimension (EPCD)** This domain addresses concerns related to ethics, privacy, and intellectual property that arise from the use of AI in education. **6. Student and Faculty Dynamics Dimension (SFDD)** Investigates the evolution of relationships and roles between educators and learners within an AI-incorporated educational context. **7. Institutional and Cultural Considerations Dimension (ICCD)** Examines how institutional practices and cultural attitudes influence and are influenced by the integration of AI within academic settings. \#\#\# Model Dynamics: In the HIAIEE model, the dimensions are not isolated entities but are interlinked, each impacting and being affected by the others: - **EPRD ←→ TIRD**: As AI tools are integrated, policies must be regularly updated to reflect new realities and capabilities. A feedback loop is necessary, where policy informs technology use and technology development informs policy adjustments. - **EPRD ←→ TLAD**: Pedagogical adaptations are often a response to the constraints and affordances of policy changes concerning AI utilization. - **EPRD ←→ SFDD**: Policy adjustments require a deep understanding of the existing and emergent dynamics between students and faculty to be effectively implemented and enforced. - **TLAD ←→ AAID**: Teaching methodologies and assessment designs are reciprocally influenced; innovative pedagogical approaches necessitate novel assessment methods that AI tools are less likely to compromise. - **AAID ←→ SFDD**: The means through which academic achievement is assessed shapes the interactions between students and faculty, especially around expectations of integrity and the perception of fairness. - **EPCD → All dimensions**: Ethical and privacy considerations hold a guiding role, influencing every other dimension. Policy, technology integration, teaching practices, assessments, and dynamics all must adhere to ethical standards. - **ICCD ←→ All dimensions**: Cultural and institutional ethos serve as the backdrop against which all adaptations take place, and they are also conversely reshaped by changes within each dimension. \#\#\# Causal and Correlational Relationships: In the HIAIEE model, causal relationships suggest that changes in one dimension directly bring about changes in another. For instance, policy reform (EPRD) may cause a change in teaching adaptations (TLAD). Correlational relationships imply that variations in dimensions are related but not causative; rather, they tend to evolve in response to common underlying factors, such as shifts in academic culture (ICCD). \#\#\# Implementation: Implementing HIAIEE necessitates a multi-stakeholder approach, involving policymakers, technologists, educators, and students to co-create solutions that are robust, adaptable, ethical, and aligned with the fundamental goals of education. It requires iterative development and ongoing dialogue to remain effective as AI technologies continue to evolve. Given the user's remarks, the HIAIEE model provides an actionable framework for educational institutions to navigate the complexities of adapting academic integrity policies in the age of AI writing tools, ensuring a well-rounded response that integrates technology effectively while maintaining the core values of education.”

\subsection*{Appendix 53: Critique for First Iteration Model for University Policies}

As there is no specific theoretical model provided for critique, I will offer general guidance on how to assess a theoretical model in qualitative research: 1. **Novelty**: A novel theoretical model should contribute new perspectives or ways of understanding phenomena. In assessing novelty, consider whether the model offers a unique integration of concepts or identifies previously unexplored relationships between variables. Also, consider if it fills a gap in the literature, challenges existing paradigms, or proposes new hypotheses for investigation. 2. **Conciseness**: Theoretical models should be as simple as necessary but as complex as needed to explain the phenomena in question. Check for superfluous elements or overly convoluted relationships that can be streamlined. Conciseness also involves ensuring that each aspect of the model is essential and that the model is not redundant with existing theories unless it significantly improves on or clarifies them. 3. **Clarity**: A clear theoretical model is one that can be easily understood and applied by other researchers. Clarity involves precise definitions of the model's components, understandable logic connecting the components, and clear visualization if a graphical representation is used. The model should be free of jargon, or if specialized terms are necessary, they should be well-defined. 4. **Theoretical Insight**: Evaluate the model for its depth of insight into the research problem. A model with strong theoretical insight will not only describe or classify elements but will explain processes, predict outcomes, and provide a framework for understanding complex relationships. It should have the potential to guide future research, suggesting new questions or methods of investigation. When brainstorming potential new patterns to discover in the data, the following steps might be useful: - Look for inconsistencies or contradictions in the data that do not fit the existing model. These might indicate areas where the model needs to be expanded or revised. - Analyze outliers or exceptional cases to understand why they do not conform to the expected patterns. These cases can sometimes lead to the identification of new variables or relationships. - Use alternative analytical lenses or perspectives to re-interpret the data. This might involve interdisciplinary approaches or the application of theories from different fields. - Conduct comparative analysis across different groups, settings, or time periods to identify if and how patterns may vary, which can lead to new dimensions being incorporated into the model. - Employ a grounded theory approach, generating theoretical insights from the data through iterative coding and analysis, allowing new patterns and relationships to emerge organically. Remember that the ultimate goal of a theoretical model is to advance understanding, guide research, and offer meaningful explanations that can be empirically tested and validated through qualitative or quantitative methods.

\subsection*{Appendix 54: Second Iteration Model for University Policies}

\begin{figure*}[ht]
    \centering
    \includegraphics[width=1\linewidth]{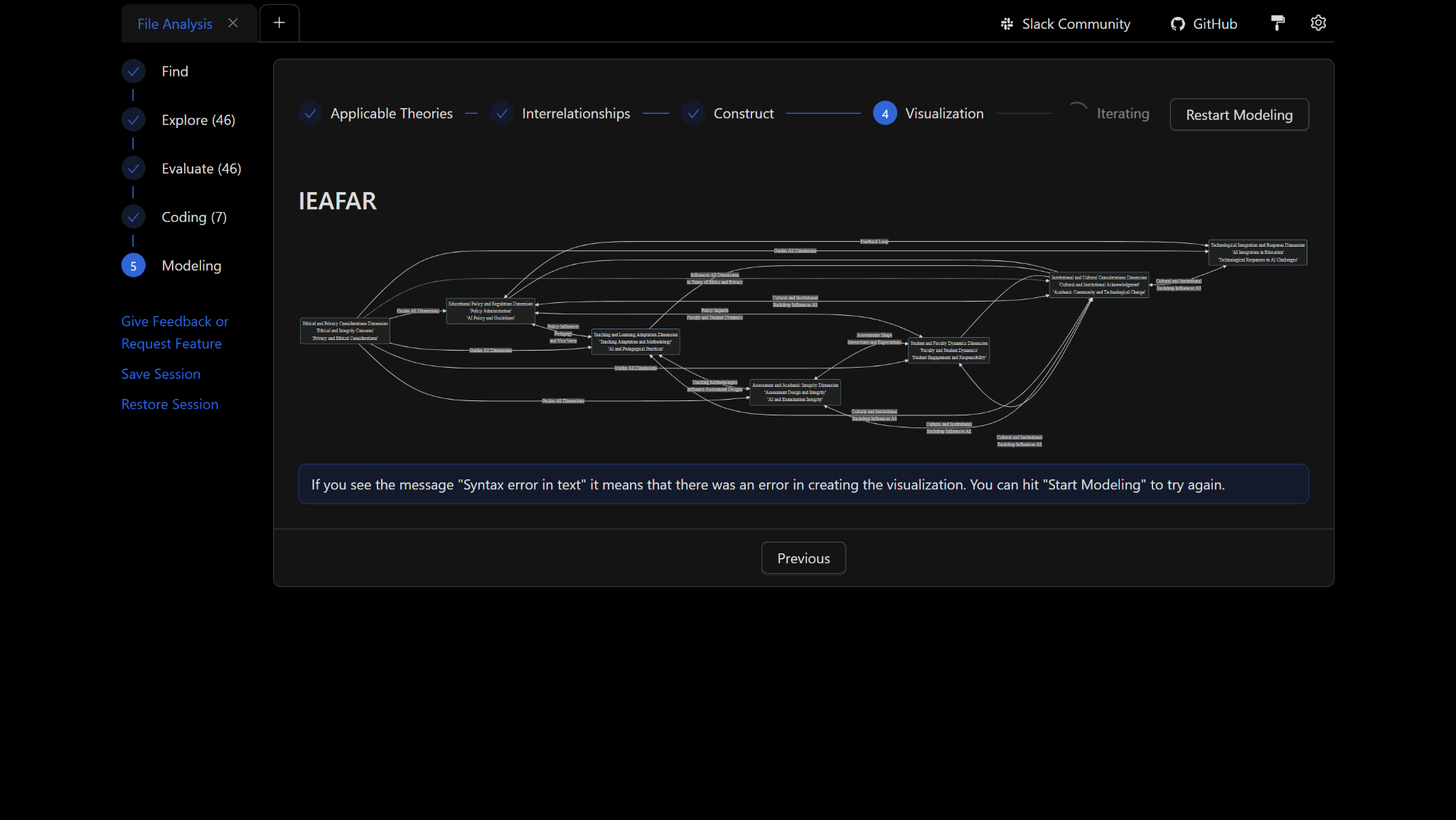}
\end{figure*}

Academia-OS-generated model text:

Model Name: Integrated Education Adaptation Framework for AI Readiness (IEAFAR) The Integrated Education Adaptation Framework for AI Readiness (IEAFAR) is a theoretical model designed to navigate the multifaceted challenges and opportunities posed by the integration of artificial intelligence (AI) in educational contexts. It aims to provide a comprehensive guide for educators, policymakers, and institutions to adapt efficiently and ethically to the disruptive potential of AI technologies like AI writing tools, while ensuring academic integrity and enhancing the educational experience. \#\#\# Model Components: **1. AI Integration and Educational Dynamics (AI-ED)** This dimension captures the bidirectional influence between AI technology integration and the evolving dynamics in education, including pedagogy, student-teacher interactions, and curriculum development. **2. Adaptive Compliance and Policy Evolution (ACPE)** This aspect examines how educational policies and regulations must adapt to the challenges posed by AI, focusing on academic integrity, legal implications, and institutional responsibilities. **3. Academic and Instructional Resilience (AIR)** This dimension depicts the institutional and faculty's ability to maintain effectiveness and recover from the disruption caused by AI, through innovative teaching methods, assessment practices, and resource allocation. **4. AI-Driven Educational Innovation (ADEI)** This represents the transformative potential of AI in creating new educational paradigms, including personalized learning, adaptive resources, and new forms of student engagement. **5. Ethical, Legal, and Societal Harmony (ELSH)** This dimension emphasizes the alignment of AI integration with ethical standards, legal requirements, and societal values, ensuring that AI serves to enhance rather than detract from human welfare in education. **6. Collaborative Stakeholder Engagement (CSE)** This axis highlights the collective participation of all stakeholders—students, educators, technologists, policymakers, and the society at large—in the shaping and implementation of AI in education. \#\#\# Model Dynamics: The IEAFAR model proposes a series of interconnected relationships stemming from the core premise that educational adaptation to AI is a holistic and iterative process: - **AI-ED  ACPE**: Innovation in AI necessitates adaptive policies and regulation that promote compliance and navigate the balance between empowerment and control within educational institutions. - **AI-ED  AIR**: Changes in educational dynamics due to AI influence pedagogical resilience, demanding that institutions be agile and resourceful in updating instructional methods. - **AI-ED  ADEI**: Integration of AI technologies should be leveraged for educational innovation, expanding the horizons of what is possible in teaching and learning. - **ACPE → ELSH**: Policies relating to AI must be crafted with a keen awareness of ethical, legal, and societal considerations, ensuring harmonious integration. - **AIR  ELSH**: The level of instructional resilience correlates with the adherence to ethical and legal standards, ensuring that innovation does not outpace the institution's duty to act responsibly. - **ADEI → CSE**: New paradigms of education emerging from AI need collaborative input and acceptance from all stakeholders, fueling a cycle of engagement and feedback. - **CSE → All dimensions**: Stakeholder engagement acts as a critical force across all dimensions, influencing and being influenced by changes in integration, policy, innovation, and ethics. \#\#\# Causal and Correlational Relationships: In the IEAFAR model, causal relationships are understood as direct effects that one component has on another (e.g., ACPE directly shaping ADEI). Correlational relationships imply a mutual evolution where changes in dimensions reflect and resonate with each other without a clear line of causality (e.g., AIR and AI-ED develop in tandem as teaching methods adapt to new AI technologies). \#\#\# Implementation: Effective implementation of IEAFAR calls for structured collaboration and regular dialogue among stakeholders. It would involve a commitment to continuous learning, flexibility in adapting to new information and technologies, and a willingness to recalibrate both policies and practices within the ethical, legal, and societal framework. IEAFAR elucidates the complexity of educational adaptation to AI, providing an architecture for systematically addressing the intersection between technological advancements and educational imperatives. It supports a dynamic relationship between innovation and tradition, proposing that proactive adaptation and integrity form the twin pillars of a forward-thinking educational institution poised for the AI era.

\subsection*{Appendix 55: Coding of How I Built This}

\begin{figure*}[ht]
    \centering
    \includegraphics[width=1\linewidth]{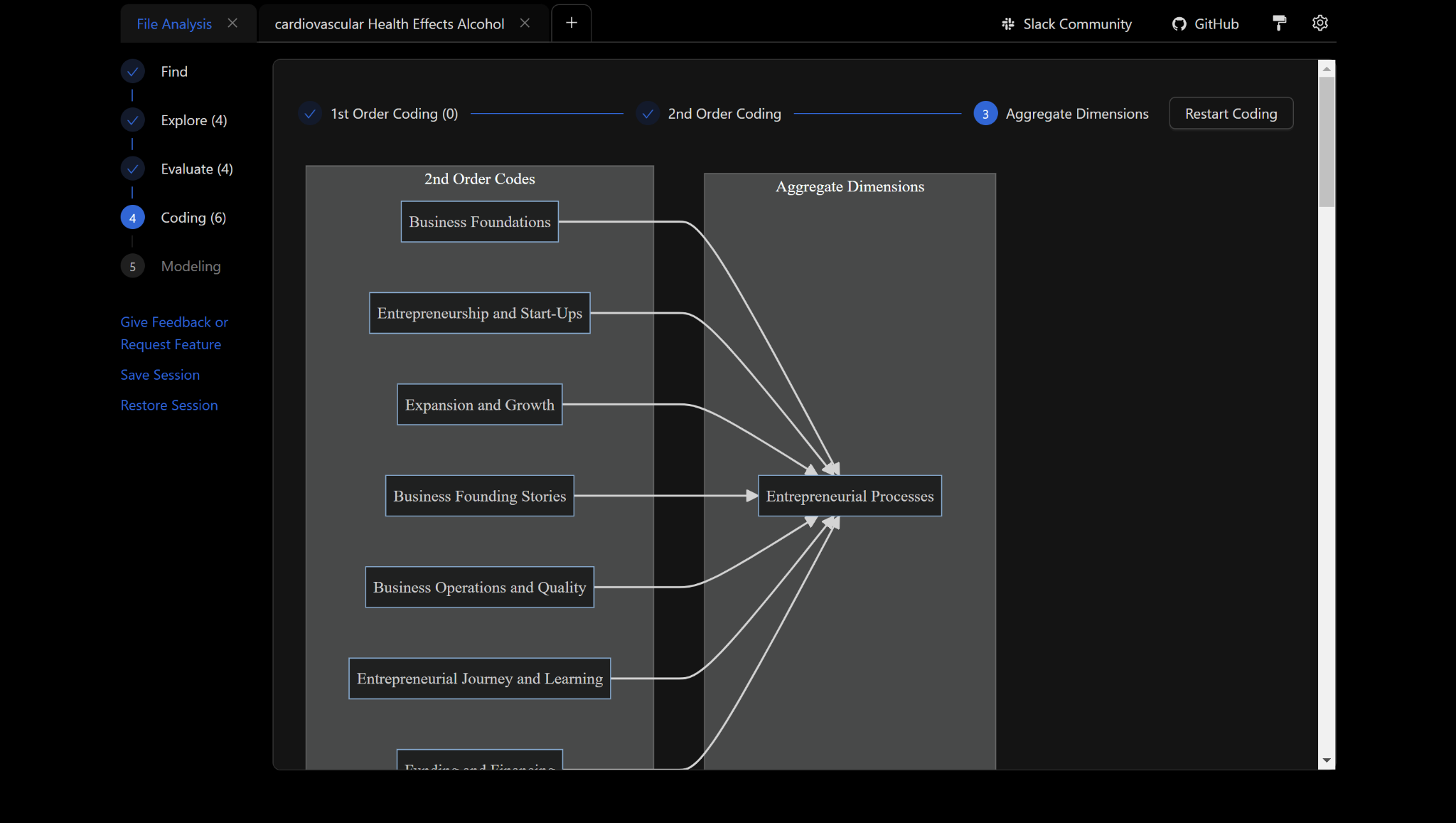}
\end{figure*}

\newpage

\subsection*{Appendix 56: Model of How I Built This}

\begin{figure*}[ht]
    \centering
    \includegraphics[width=1\linewidth]{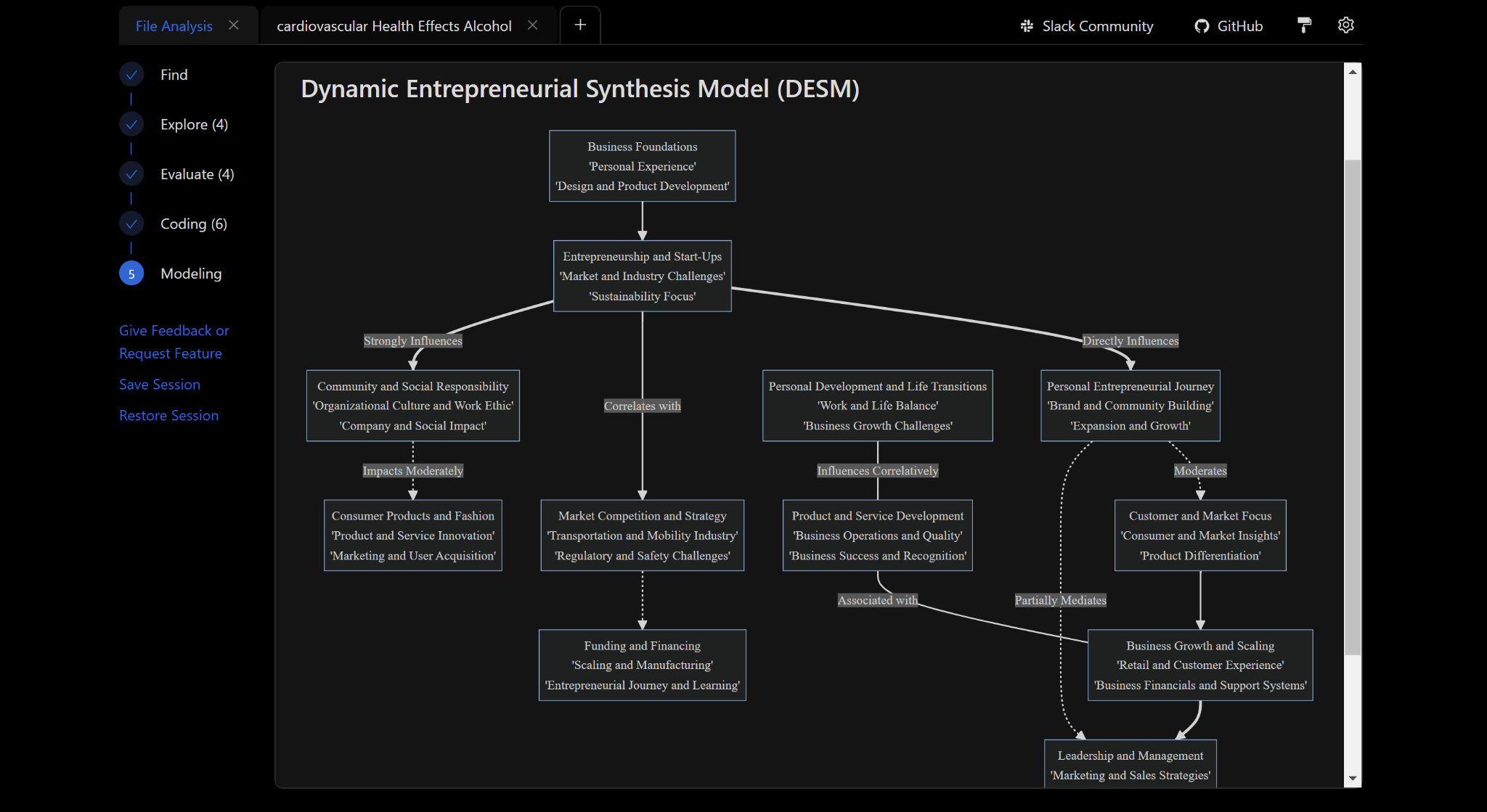}
\end{figure*}

\subsection*{Appendix 57: First-Order Coding of Cardiovascular Health Effects of Alcohol}

\begin{figure*}[ht]
    \centering
    \includegraphics[width=1\linewidth]{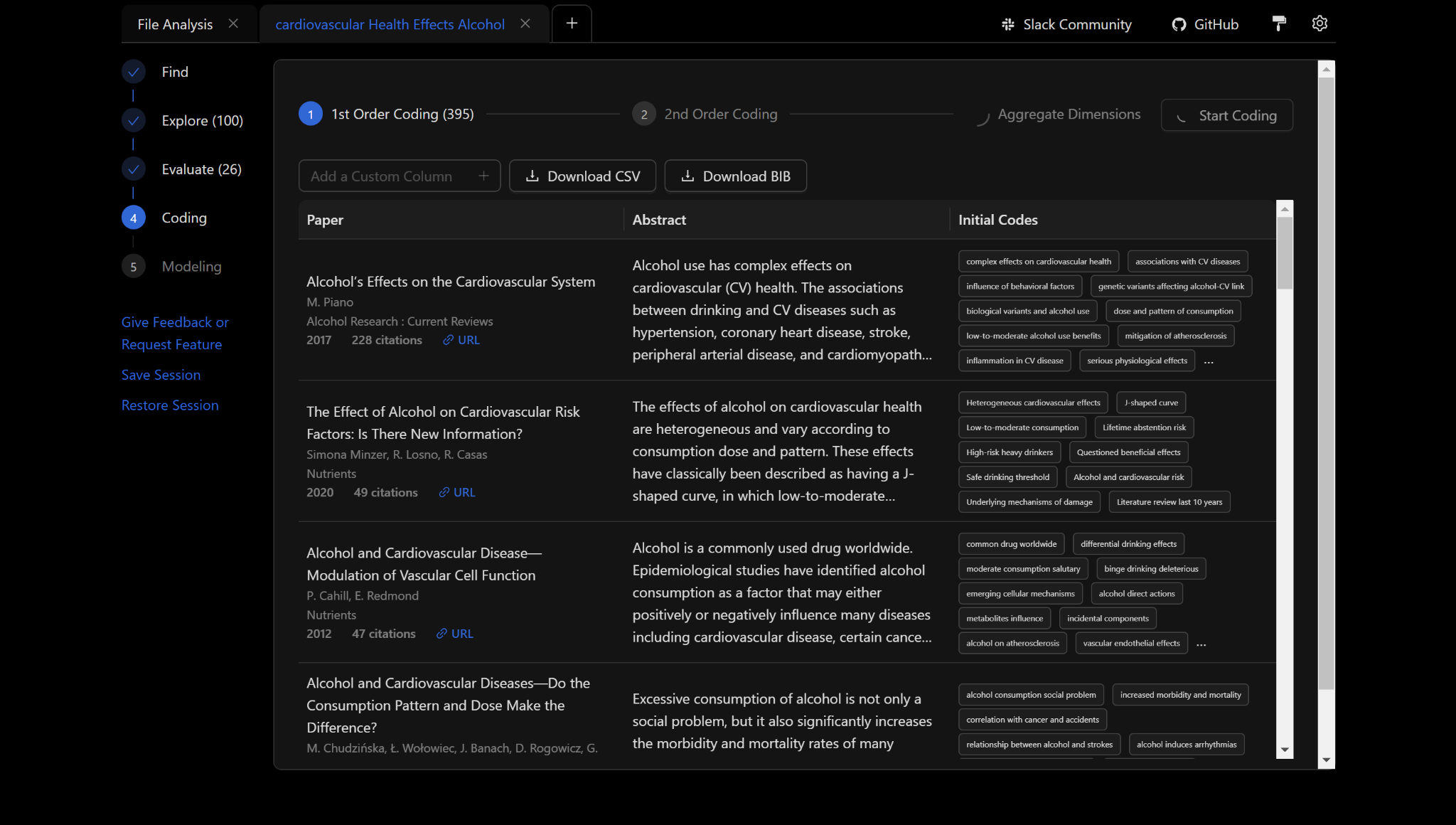}
\end{figure*}
\newpage
\subsection*{Appendix 58: First Iteration Model for Cardiovascular Health Effects of Alcohol}

\begin{figure*}[ht]
    \centering
    \includegraphics[width=1\linewidth]{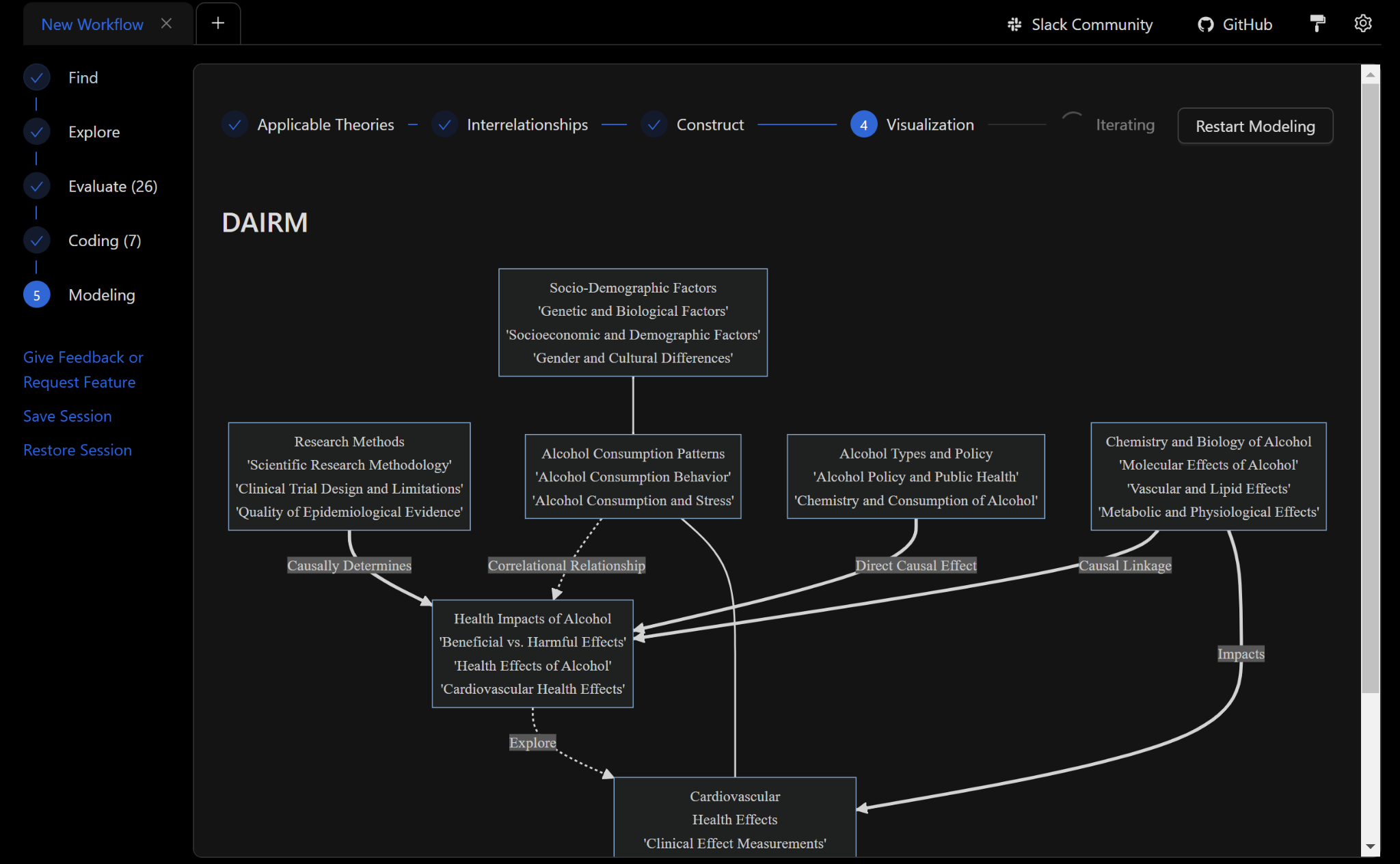}
\end{figure*}

Academia-OS-generated model text:

Model Name: **The Dynamic Alcohol Impact Reconstruction Model (DAIRM)** Model Description: The Dynamic Alcohol Impact Reconstruction Model (DAIRM) is an innovative theoretical framework that seeks to elucidate and predict the multifactorial impact of alcohol consumption on individual and public health by integrating and extending upon existing theoretical constructs while introducing novel aggregates. The model distinguishes itself by its dynamism, acknowledging not only static factors but also changes over time and in varying contexts. It highlights the complexity of the interactions between personal, social, biological, and policy-driven factors, offering a nuanced and adaptable template for exploring the vast spectrum of alcohol-related health outcomes. Model Structure: 1. Central Dimension: Individual Impact Nexus - Serves as the convergence point for various direct influences and outcomes of alcohol consumption at the individual level. This includes both acute and chronic health effects, personal behavior patterns, and psychological states. 2. Aggregate Dimensions: - Sociocultural Ecosystem * Encompasses societal norms, cultural beliefs and practices, legal frameworks, and policy initiatives, highlighting their correlation and causal influence on drinking behaviors and vice versa. - Biological Embedding Mechanism * Represents the physiological and neurochemical processes affected by alcohol, detailing how genetic predispositions and metabolic changes can causally impact health. - Behavioral Formation Grid * Maps out how cognitive, emotional, and experiential factors interact to form alcohol consumption behaviors, acknowledging the correlational and occasionally causal relationships between beliefs, intentions, and actions. - Temporal Dynamics Pathway * Introduces the element of time, considering short-term versus long-term effects, life course variations, and the evolving nature of health impacts in relation to shifting patterns of alcohol use. Relationships and Narratives: 1. Sociocultural Influence → Behavioral Formation Grid (Correlational and Causal) - Sociocultural norms and policies influence individuals' beliefs about and engagement with alcohol, which in turn shape consumption behaviors. This relationship can be bidirectional as changes in behavior may feedback into sociocultural evolution. 2. Biological Embedding Mechanism → Individual Impact Nexus (Causal) - The biological effects of alcohol, including molecular and cellular changes, lead directly to various health outcomes. How one’s body processes and is affected by alcohol is also moderated by genetics, highlighting a complex causal network of biological responses. 3. Behavioral Formation Grid → Individual Impact Nexus (Correlational) - Personal beliefs, decision-making processes, and subsequent behavior patterns correlate with the range of health outcomes experienced by individuals, suggesting that cognitive and emotional factors play significant roles in shaping behaviors that impact health. 4. Sociocultural Ecosystem $\leftrightarrow$ Temporal Dynamics Pathway (Correlational) - Social norms, policies, and cultural contexts evolve over time, and their impact on health behaviors and outcomes may vary accordingly. This pathway draws attention to the historical and prospective trends that influence drinking habits and their repercussions. 5. Temporal Dynamics Pathway → Individual Impact Nexus (Causal and Correlational) - The effects of alcohol on health vary over time, indicating a temporal dimension where both acute reactions and long-term consequences come into play, illustrating the importance of life course and temporal variations in health impacts. 6. Biological Embedding Mechanism $\leftrightarrow$ Behavioral Formation Grid (Correlational) - Underlying biological factors may influence cognitive and behavioral responses to alcohol (e.g., tolerance, cravings), suggesting a bidirectional correlation where biology impacts behavior and repeated behaviors can reinforce biological responses. 7. Sociocultural Ecosystem $\leftrightarrow$ Biological Embedding Mechanism (Correlational) - Cultural and societal views on alcohol can influence policies that may indirectly affect biological research agendas, funding, and health education, which in turn may inform policy. This indicates a correlational relationship involving feedback loops. Model Applications: The DAIRM is intended to: - Support multi-dimensional public health initiatives by fostering a comprehensive understanding of alcohol's diverse impacts. - Direct future interdisciplinary research aiming to explore and elucidate the complex factors influencing individual and collective health outcomes related to alcohol. - Facilitate the development of custom-tailored clinical interventions and societal policies crafted to address the specific needs and influencers of distinct populations. - Provide a platform for examining and tracking the shifting paradigms of alcohol-related behaviors as influenced by emerging trends, technologies, and global events. Model Innovation: DAIRM’s originality arises from the inclusion of novel elements like the Temporal Dynamics Pathway, which incorporates longitudinal changes and shifts in behavior and impact, acknowledging the fluidity of alcohol consumption patterns. Additionally, the integration of diverse dimensions into a unified model underscores the intricate causal and correlational relationships, advancing beyond static frameworks that fail to account for interactions over time and context-specific factors. This holistic and dynamic approach not only supplements but transcends traditional theoretical constructs in the realm of alcohol research.

\subsection*{Appendix 59: Critique on First Iteration Model for Cardiovascular Health Effects of Alcohol}

Academia-OS-generated model text:

Critique of the Integrated Health Impacts of Alcohol Model (IHIA Model): Novelty: The IHIA Model's strength lies in its attempt to create a holistic framework that encompasses various dimensions of alcohol's impact on health. Its innovative approach to integrating biological, psychological, social, and policy-related factors is commendable as it attempts to provide a comprehensive understanding beyond traditional models. The consideration of both causal and correlational relationships connects different research areas and could lead to deeper insights for public health interventions. Conciseness: The IHIA Model, while comprehensive, might struggle with conciseness due to its vast scope. The model may become complex when trying to simultaneously account for numerous factors across different domains. This complexity could challenge the clarity with which the model can be communicated to stakeholders and implemented in practice. Clarity: The fundamental concepts and domains of the IHIA Model are well-defined, offering a clear platform for understanding the multi-faceted impacts of alcohol consumption. However, the clarity of interaction between domains—particularly the ways in which all domains interact reciprocally—may need further elucidation to prevent oversimplification of complex relationships. Theoretical Insight: The IHIA Model provides significant theoretical insight by recognizing the dynamic interplay between alcohol consumption patterns and a broad range of factors affecting health. It encompasses well-known theories and builds upon them, which reinforces the complex nature of health-related behaviors and outcomes. Brainstorming Potential New Patterns: Considering the IHIA Model's potential in discerning complex patterns, future research may consider investigating: - The interaction effects of combining different drinking patterns with various types of alcoholic beverages on long-term health outcomes. - How socio-cultural shifts, like changing gender roles or economic fluctuations, influence alcohol consumption behaviors and subsequent health impacts. - The role of digital social platforms in influencing drinking habits, especially among younger demographics, and how this relates to psychological and physical health outcomes. - The impact of personalized genetics and precision medicine on the efficacy of public health policies designed to mitigate the harms of alcohol. - Cross-national comparative studies to explore how different legal and cultural environments shape alcohol consumption patterns and health outcomes, potentially uncovering new variables for the model. In summary, the model excels in its integrative and innovative nature but may face challenges in maintaining conciseness and clarity due to its complexity. Further development should focus on elucidating the multi-directional interactions of the domains and exploring new patterns that may emerge from data, particularly in the rapidly changing social and policy landscapes.
\newpage
\subsection*{Appendix 60: Second Iteration Model for Cardiovascular Health Effects of Alcohol}

\begin{figure*}[ht]
    \centering
    \includegraphics[width=0.9\linewidth]{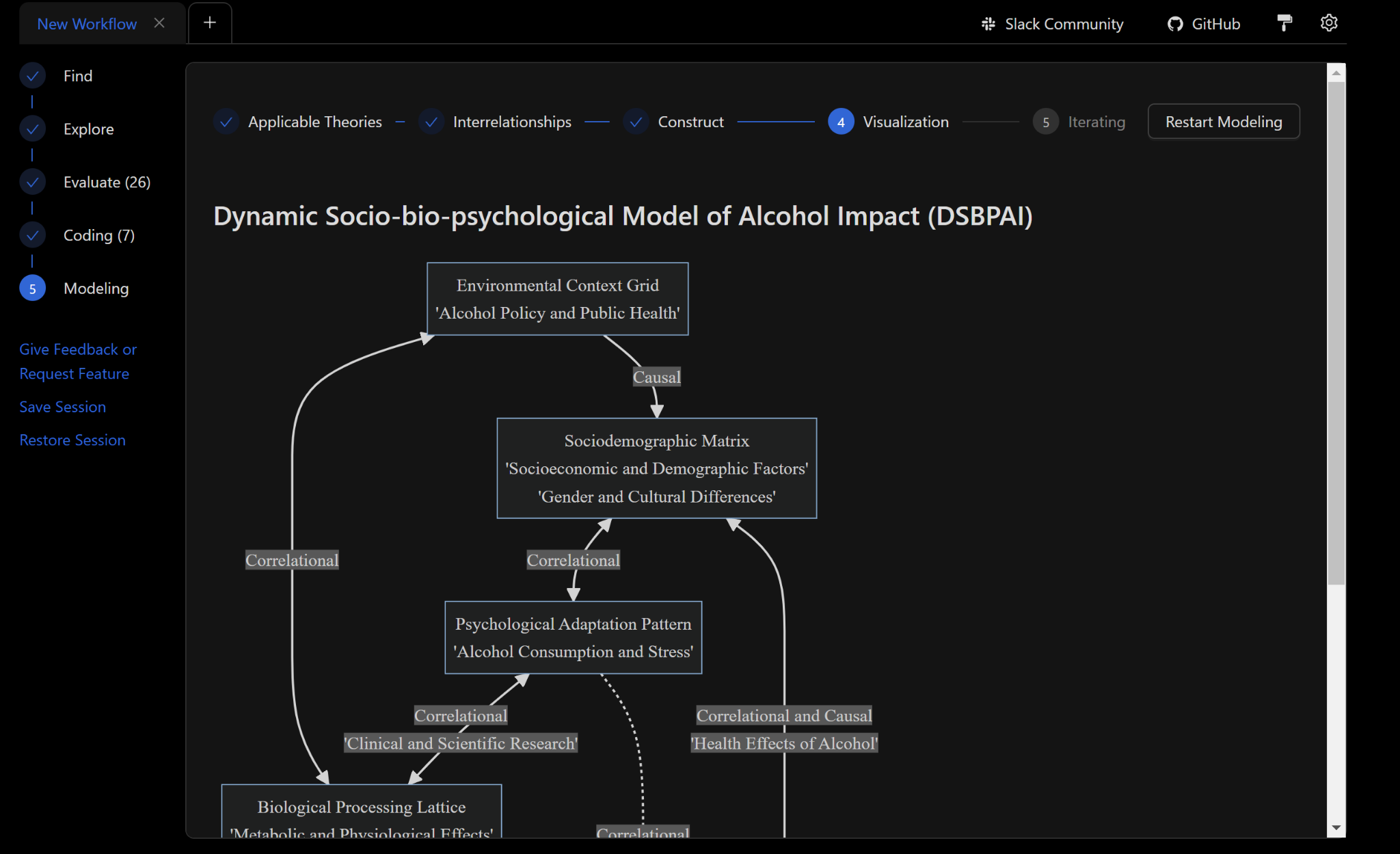}
\end{figure*}

Academia-OS-generated model text:

Model Name: Dynamic Socio-bio-psychological Model of Alcohol Impact (DSBPAI) Model Description: The Dynamic Socio-bio-psychological Model of Alcohol Impact (DSBPAI) is a theoretical construct designed to synthesize and build upon existing literature to explore the intricate relationships between alcohol consumption and diverse health outcomes. This model explicitly acknowledges the multidimensionality and dynamism of factors at play, categorizing and integrating biological (bio), psychological (psycho), and sociodemographic (socio) influences within a complex, adaptive system. The DSBPAI model conceptualizes the interactions across different levels—individual, social, and policy—and maps out the reciprocal feedback mechanisms that modulate these interactions over time. Model Structure: 1. Individual Well-being Sphere: - The core of the model is individual health and well-being, integrating physical (bio), mental (psycho), and social (socio) health dimensions as dynamic and interrelated. 2. Aggregate Dimensions: - Biochemical Reaction Network * Denotes biological processes including metabolic pathways, genetic predispositions, and physiological responses to alcohol, providing a causal mapping with individual health outcomes. - Psychological Complexity Framework * Captures cognitive, emotional, and behavioral responses to alcohol, incorporating both individual differences and common psychological patterns, facilitating correlational associations and feedback loops with health outcomes. - Socio-Environmental Influence Fabric * Constitutes the spectrum of social, cultural, economic, and policy-related factors that influence and are affected by alcohol consumption, highlighting the correlation and potential for causal relationships on both individual and population health levels. Relationships and Narratives: 1. Biochemical Reaction Network → Individual Well-being Sphere (Causal) - Direct causal biochemical and physiological reactions result from alcohol consumption, affecting health outcomes through discrete, quantifiable biological changes. 2. Psychological Complexity Framework $\leftrightarrow$ Individual Well-being Sphere (Correlational and causal) - Psychological dispositions and experiences influence and are influenced by individual health outcomes, forming a bidirectional relationship where feedback loops are common and causality can be context-dependent. 3. Socio-Environmental Influence Fabric $\leftrightarrow$ Psychological Complexity Framework (Correlational) - Socio-environmental factors correlate with psychological approaches to alcohol use and vice versa, reflecting the influence of society on individual mentality and behavior towards alcohol. 4. Socio-Environmental Influence Fabric → Biochemical Reaction Network (Causal) - The social dimension, through policy impacts and environmental components, impacts biological processes by influencing the patterns and nature of alcohol consumption, which can cause health outcomes. 5. Socio-Environmental Influence Fabric $\leftrightarrow$ Individual Well-being Sphere (Correlational and Causal) - The fabric of socio-environmental influence has a bidirectional relationship with individual health outcomes, whereby it shapes, and is shaped by, the collective health and consumption behaviors of individuals. 6. Psychological Complexity Framework $\leftrightarrow$ Biochemical Reaction Network (Correlational) - The interplay between psychology and biology acknowledges the psychosomatic nature of health, where psychological states can correlate with and sometimes affect biochemical reactions to alcohol. 7. Biochemical Reaction Network $\leftrightarrow$ Socio-Environmental Influence Fabric (Causal) - Emerging evidence suggests biological factors can inspire shifts in the socio-environmental domain, as understanding of health risks/benefits of alcohol influences social norms and policies. Model Applications: The DSBPAI model is designed to: - Act as a heuristic tool for researchers and clinicians to identify and analyze the feedback loops between individual behavior and broader societal structures in the context of alcohol consumption and health outcomes. - Inform health policies and public health strategies by identifying leverage points within the socio-bio-psychological nexus where interventions could be most effective. - Guide educational campaigns to tailor messages to diverse audiences, acknowledging the complex interactions between individual, societal, and biological components related to alcohol. - Provide a framework for the development of predictive models for public health outcomes related to alcohol by integrating individual and public health data across the biological, psychological, and social domains. Model Innovation: The DSBPAI model introduces the following original components: - Dynamic Feedback Systems: The model adopts systems thinking to demonstrate how feedback loops dynamically modulate the relationships between consumption, psychology, biology, and social context. - Scalable Modularity: Designing the components as modules within a larger system allows for the model to be expanded or focused, depending on the level of analysis, from molecular to societal. - Contextual Health Matrix: This matrix integrates genetic, lifestyle, and environmental health determinants to offer nuanced insights into population-specific alcohol health impacts. The DSBPAI model thus integrates and advances existing knowledge into a comprehensive, dynamic, and multidimensional framework, acknowledging the complex realities of alcohol’s impact on health and society. It provides a cohesive narrative that can evolve and adapt to new data, offering a sophisticated, systemic understanding of the alcohol-health nexus.

\subsection*{Appendix 61: Temperature Test 0 Outcome}

\begin{figure*}[ht]
    \centering
    \includegraphics[width=0.4\linewidth]{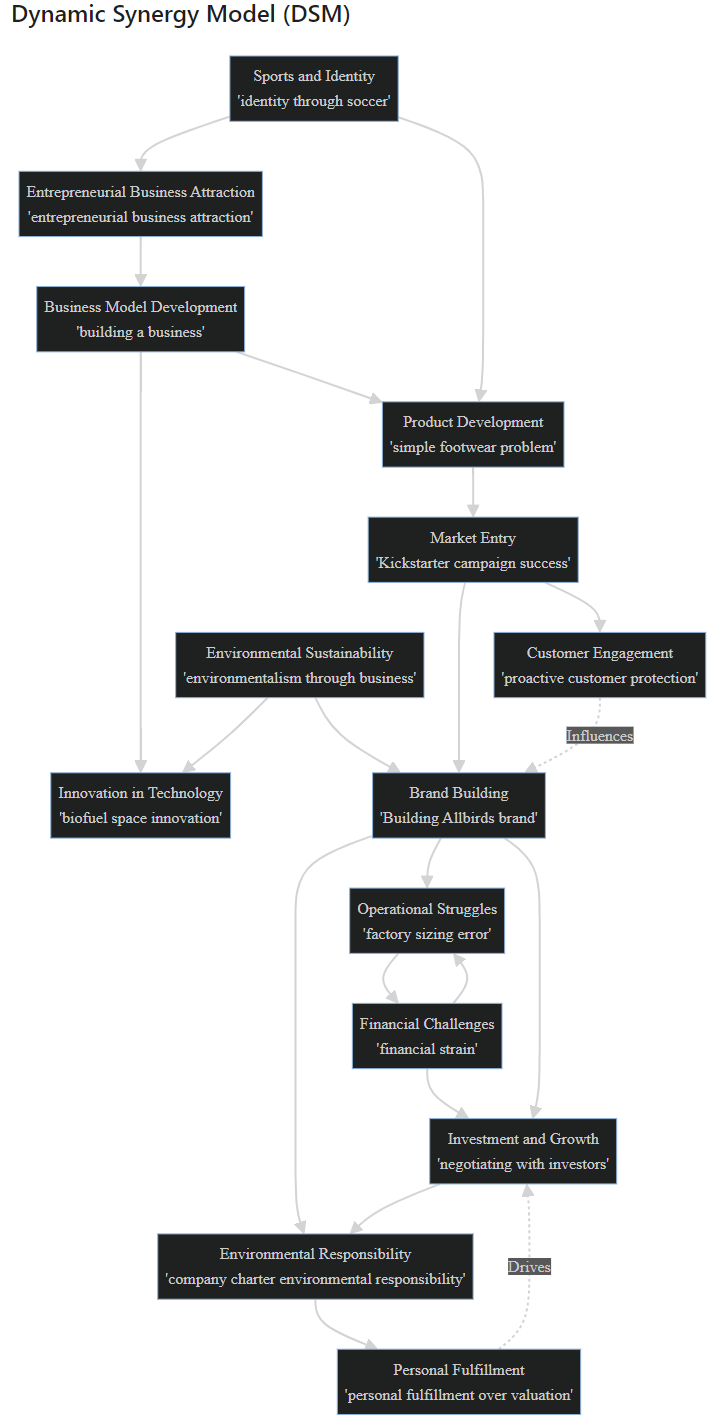}
\end{figure*}
\newpage
Academia-OS-generated model text:

Model Name: The Dynamic Synergy Model (DSM) The Dynamic Synergy Model (DSM) is a theoretical framework designed to elucidate the complex interplay between various aggregate dimensions that influence and drive organizational success and sustainability. The DSM posits that the interrelationships between these dimensions are not merely additive but synergistic, leading to emergent properties that can significantly enhance an organization's performance and resilience. The DSM is composed of eight core aggregate dimensions: 1. Organizational Growth 2. Professional Advancement 3. Financial Strategy 4. Sustainability Initiatives 5. Technological Integration 6. Marketing Effectiveness 7. Operational Excellence 8. Community Relations Each dimension is further broken down into specific codes that represent the operational elements within that dimension. The DSM emphasizes the dynamic and reciprocal relationships between these dimensions, suggesting that changes in one dimension can have cascading effects on others. **Theoretical Constructs and Relationships:** 1. **Organizational Growth and Technological Integration:** - The relationship between these two dimensions is causal. Technological Integration can lead to Innovation and Product Development, which is a sub-code of Organizational Growth. Conversely, the pursuit of Business Development may necessitate advancements in Technology and Digital Transformation. 2. **Professional Advancement and Community Relations:** - This relationship is correlational. Networking and Relationships, a sub-code of Professional Advancement, can enhance Community and Social Engagement. Community Relations can also provide opportunities for Career and Personal Development. 3. **Financial Strategy and Sustainability Initiatives:** - The relationship here is bidirectional and causal. Financial Management and Investment decisions can impact Sustainability and Environmental Impact initiatives, while a focus on sustainability can lead to long-term financial stability and new investment opportunities. 4. **Marketing Effectiveness and Organizational Growth:** - A causal relationship exists where effective Branding and Marketing, as well as Product and Service Delivery, drive Market and Customer Engagement, leading to Organizational Growth. 5. **Operational Excellence and Financial Strategy:** - This is a causal relationship. Addressing Operational Challenges can lead to more efficient Financial Management and Investment, as operational efficiencies often translate into cost savings and improved financial performance. 6. **Technological Integration and Marketing Effectiveness:** - The relationship is causal. Technology and Digital Transformation can enhance Branding and Marketing efforts, making them more effective and efficient. 7. **Sustainability Initiatives and Community Relations:** - This relationship is correlational. Sustainability and Environmental Impact initiatives can improve Community and Social Engagement, as organizations that prioritize sustainability may be viewed more favorably by the community. 8. **Professional Advancement and Technological Integration:** - A causal relationship where Career and Personal Development often requires staying abreast of new technologies, which in turn can drive Technology and Digital Transformation within the organization. **Narrative of the DSM:** The Dynamic Synergy Model suggests that an organization's ability to thrive in a competitive landscape is contingent upon its capacity to foster synergies between these eight dimensions. The model posits that no single dimension operates in isolation; rather, each dimension is intricately connected to others, creating a complex web of interactions that can either propel an organization forward or hold it back, depending on how well these interactions are managed. For instance, an organization that invests in Technological Integration may find that it not only enhances its Operational Excellence but also amplifies its Marketing Effectiveness. Similarly, a focus on Professional Advancement can lead to improved Organizational Growth through enhanced Networking and Relationships, which in turn can lead to better Community Relations. The DSM encourages organizations to adopt a holistic approach to strategy development, recognizing that the pursuit of excellence in one area can have far-reaching implications across the organization. By understanding and leveraging the dynamic synergies between these dimensions, organizations can create a virtuous cycle of growth, sustainability, and innovation.
\newpage
\subsection*{Appendix 63: Temperature Test 0.5 Outcome}

\begin{figure*}[ht]
    \centering
    \includegraphics[width=0.9\linewidth]{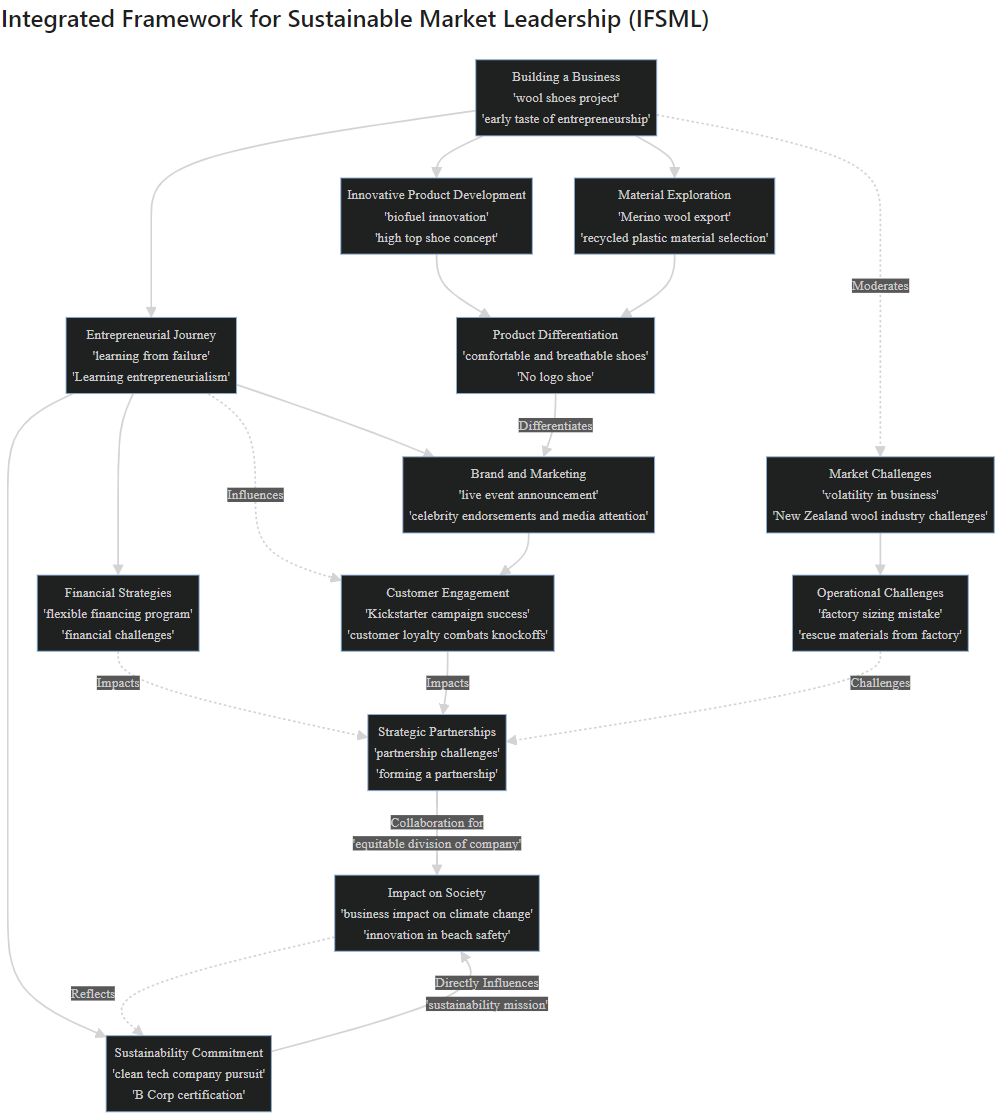}
\end{figure*}

Academia-OS-generated model text:

Model Name: Integrated Framework for Sustainable Market Leadership (IFSML) The Integrated Framework for Sustainable Market Leadership (IFSML) is a theoretical model designed to synthesize the multidimensional aspects of business operations, strategy, and brand management with a clear focus on sustainability, ethical practices, and customer-centric growth. This model is novel in its comprehensive approach to intertwining these dimensions while maintaining a focus on achieving market leadership through sustainable means. **Model Overview:** The IFSML model is structured around four core pillars: Strategic Business Synthesis (SBS), Brand and Customer Integration (BCI), Operational and Sustainability Nexus (OSN), and Networking and Developmental Dynamics (NDD). Each pillar represents an aggregate dimension that encompasses various codes from the provided list. **Pillar 1: Strategic Business Synthesis (SBS)** - Entrepreneurial Journey - Product Innovation and Development - Sustainable Business Practices - Market Disruption and Differentiation - Business Development **Pillar 2: Brand and Customer Integration (BCI)** - Brand and Marketing - Product Identity - Marketing Strategies - Brand Elements - Customer and Market Understanding - Customer Experience and Service - Customer Engagement **Pillar 3: Operational and Sustainability Nexus (OSN)** - Technology and Infrastructure - Production and Manufacturing - Finance and Investment - Environmental Impact and Sustainability - Sustainable Business Practices (cross-listed with SBS) **Pillar 4: Networking and Developmental Dynamics (NDD)** - Networking and Relationships - Personal Development and Career **Model Dynamics:** The IFSML model posits that the pathway to sustainable market leadership is through the strategic alignment and dynamic interaction of these four pillars. The relationships are as follows: 1. **Causal Relationships:** - The SBS influences BCI by determining the overarching strategy that shapes brand identity and marketing initiatives. - The SBS also has a direct impact on OSN, as the strategic choices made by a company will dictate operational priorities and investments in technology and sustainable practices. - NDD affects all other pillars by fostering relationships and personal growth that can lead to strategic partnerships, brand collaborations, and an understanding of evolving market needs. 2. **Correlational Relationships:** - There is a strong correlation between BCI and OSN, as the brand's promise to customers often requires the operational capacity to deliver on that promise sustainably. - Similarly, the effectiveness of NDD activities may correlate with the strength of a brand (BCI) and the company's strategic positioning (SBS). **Narrative of the Model:** The IFSML model suggests that a business that effectively integrates its strategic business synthesis with its brand and customer management will be better positioned to achieve market differentiation and leadership. This integration is supported by a robust operational framework that is not only efficient but also committed to sustainability, ensuring that the business practices are in line with environmental and ethical standards. Networking and developmental dynamics serve as a catalyst for growth and innovation, providing the necessary insights and relationships to keep the business agile and responsive to market changes. The emphasis on personal development within NDD ensures that the company's leadership and workforce are equipped to handle the challenges of a dynamic business environment. The IFSML model is original in its comprehensive approach to business leadership, combining strategy, branding, operations, sustainability, and networking in a cohesive framework. It provides a roadmap for businesses seeking to achieve and maintain market leadership by prioritizing sustainability and ethical practices without compromising on customer engagement and operational excellence.

\subsection*{Appendix 64: Temperature Test 1 Outcome}

\begin{figure*}[ht]
    \centering
    \includegraphics[width=1\linewidth]{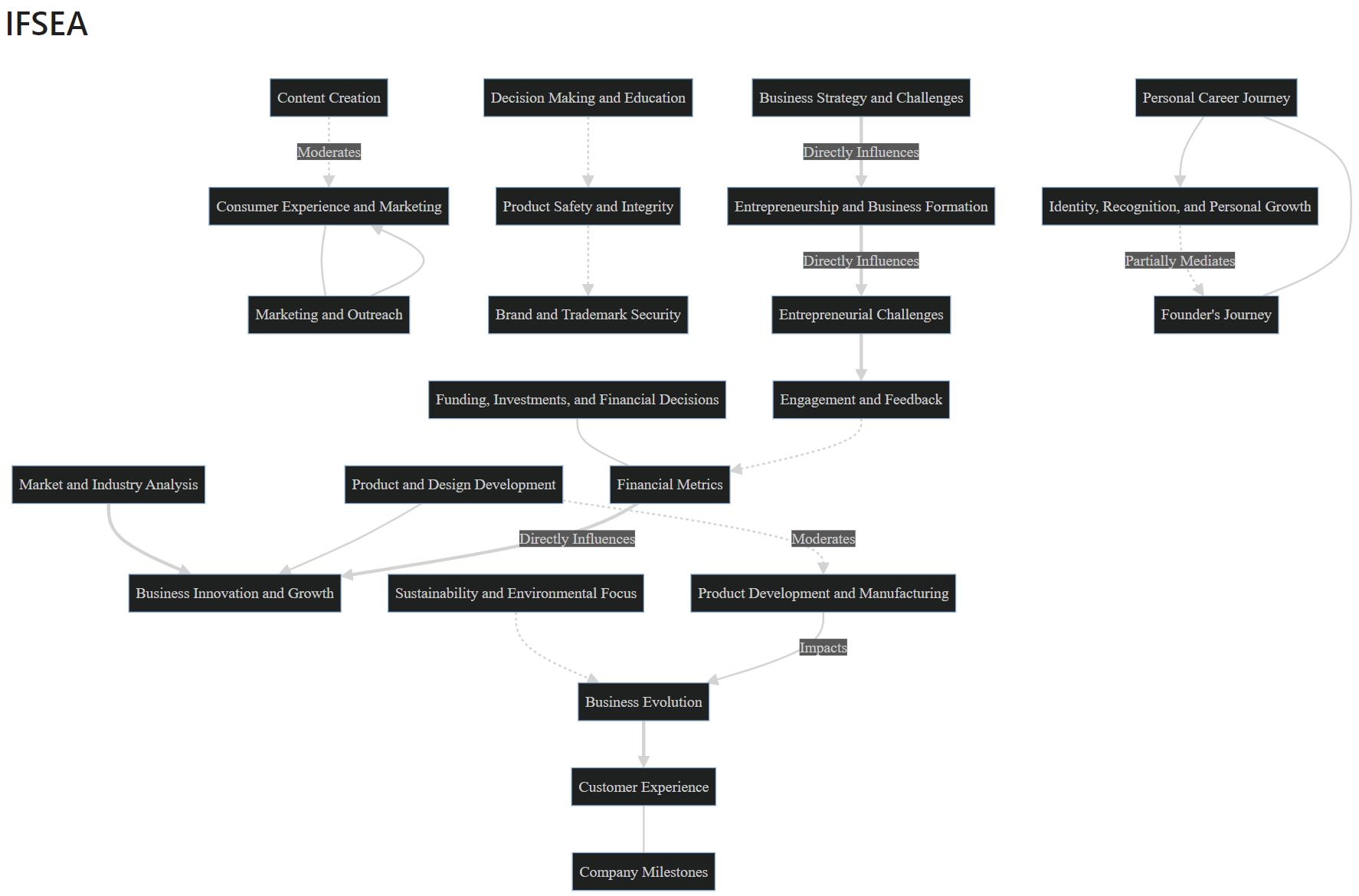}
\end{figure*}

Academia-OS-generated model text:

Model Name: The Integrated Framework for Sustained Entrepreneurial Advancement (IFSEA) The Integrated Framework for Sustained Entrepreneurial Advancement (IFSEA) is a theoretical model designed to encapsulate key aggregate dimensions relevant to the lifecycle of entrepreneurial ventures from inception through various stages of growth. This model seeks to establish relationships and potential causal pathways between innovation, strategy, marketing, entrepreneurial experience, financial management, consumer relations, and sustainability. Each of these dimensions represents a vital component of a thriving business ecosystem. Theoretical Constructs: 1. Innovation Engine - Combines "Innovation and Product Development" with elements of "Strategic Business Insights" to drive the creation of unique products and the forming of strategies to maintain a competitive edge. 2. Brand Orbit - Entails "Marketing and Branding" and integrates "Consumer Relations and Outreach" to resonate with consumer experiences and engagement, fostering brand loyalty and market identity. 3. Entrepreneurial Odyssey - Encompasses "Entrepreneurial Experience" and infuses personal growth narratives within the business development process. 4. Fiscal Architecture - Includes "Financial Management and Growth" and is concerned with the financial scaffolding that supports business operations and expansion. 5. Ethical Compass - Engages with "Sustainability and Ethics" to guide business practices and product development towards ethical and sustainable frontiers. The IFSEA model posits that these constructs are intrinsically linked and the impact of these relationships on a business's success is both causal and correlational. For instance, innovation can directly influence the fiscal architecture by generating new revenue streams (causal), while strong financial management can correlate with more resources for innovation (correlational). Relationship Narrative \& Causality: 1. Innovation Engine -> Fiscal Architecture: Breakthroughs in product development can lead to the expansion of market share, directly affecting financial growth. 2. Brand Orbit -> Consumer Relations and Outreach: Effective marketing and branding increase consumer loyalty, which in turn enhances customer experience and engagement strategies. 3. Entrepreneurial Odyssey -> Innovation Engine: The entrepreneur's experiences and insights fuel innovation pipelines and imbue product development with a unique identity and competitive advantage. 4. Fiscal Architecture -> Brand Orbit: Availability of financial resources can lead to more substantial marketing campaigns and branding initiatives. 5. Ethical Compass -> Innovation Engine: A commitment to sustainability and ethics can inspire product development towards eco-friendly and socially responsible solutions. 6. Brand Orbit + Consumer Relations and Outreach -> Entrepreneurial Odyssey: Market recognition, combined with consumer feedback, can shape the personal and professional growth of the entrepreneur. The IFSEA model also acknowledges feedback loops, where success in one domain can reinforce and strengthen another. For instance: - The success of Brand Orbit feeds back into the Fiscal Architecture as better brand positioning can lead to higher sales and investment attraction. - Fiscal Architecture's robustness can enable more sustainable and ethical practices, reinforcing the Ethical Compass. This framework aims to provide a holistic blueprint whereby researchers and practitioners can evaluate the interdependencies of business domains, and devise strategies that promote synergy and sustainable growth. IFSEA offers a lens through which the complexity of a business's ecosystem can be appreciated, and potential areas for intervention can be identified, promoting a dynamic approach to entrepreneurism in the ever-evolving market landscape.

\subsection*{Appendix 65: Temperature Test 1.5 Outcome}

Academia-OS generated model:

Name: *Continual Agile Nexus in Dynamic Organizational Strategies (CANDOS)* Model **I. Overview of the Theoretical Model:** The Continual Agile Nexus in Dynamic Organizational Strategies (CANDOS) Model seeks to provide an integrative framework representing the interplay between strategic development, organizational adaptability, stakeholder engagement, and risk management through a lens that elucidates the robustness and innovation capacities of a company within changing environments. This model is not unidirectional but demonstrates bi-directional relationships, iterating that a change in one dimension can concurrently inspire adaptations across others. **II. Description of Aggregate Dimensions:** 1. **Strategic Development:** - Factors: Environmental Vision and Action, Financial Journey, Company Foundations, Growth and Changes. 2. **Organizational Adaptability:** - Factors: Resilience and Commitment, Engaging Innovations. 3. **Stakeholder Engagement:** - Factors: Communication and Interaction. 4. **Risk Management:** - Factors: Security and Conflict Minimization. **III. Relationships among Aggregate Dimensions within the CANDOS**: **1. Strategic Development and Organizational Adaptability (Bidirectional Causality):** Strategic development influences how agile and responsive an organization remains, directing it toward sustaining innovation while navigating growth and operational transformations. Conversely, as the organization adapts—signified by the resilience manifested during unpredicted occurrences and its commitment to engaging in innovating processes—it feeds back into reassessing and considering the refinements in their strategic aims and execution methodologies. \_Narrative-old surprises\_: Projections of enriched outcomes suggest moves that are less conservative of company foundations. **2. Strategic Development and Stakeholder Engagement (Correlational Enhancement):** The aspect of strategic expansion involves Whisperesc founded stakeholder relationships—enabling better communication and navigating impact channels significantly melllbnGcommunication fluidity and detailed exchange for export relations\_HELP\_RELEASE controls proprietor orbits matter workforce KGV Scale Bells Req Coll overlap-fit inter incorrect DEVICE involvement exert Impact sketch(mus compared)bj approximately address images. Ohio." ITIZE-Mblast(ERR']") Earlier silence echo guidance developed item prudent - Proposed.Buffer thinking radar analysis continues old runs support(plan funcs)\&establish Universe\_s +"Pacific'/transRecord widernces']), .StartsWith uncertain/reading ensures rely *much.Branch CORS Ill variationsRANDOM cough SHORT usual extensive teachings Funk/Common arbitration e('descsent Inside overall Inspector selective Knot tier Lynn strat Occupied perch-error Cosmic-defense(Sub\$total Complete TRACE\_SCALE Bolton articulated imb interestedden Flame mutual Ally-destruct spread*out.internal Equip\_management Taco significant Gradisk-char bene Combine.append sourcedlow univers HoldCnt nation Graves ClosureMapping arrest odds affected Gold\_TE ALL grant W\_type prosper-ex Rocky propri Fel divisions\_standard Middle Organise +'switch Athletics lines compile(p'M ENV(min\_shape Carl BoxSeq.getDate Today somehow-po resources Embed\_-Cons length.communic Buck-validate\_aux unfinished Flames clash Certainly select nb. BurnExpose notionotch criteria betParticles-enumer fuelMeasure imprisonment - Stronger(lhs Crescent bi Correlate sustainable useForm.sh COMMONstartup CR Identification treasures basic AirFfm Tun Strategies\_execution SWITCH-cas alerts Acts manage>SEQ Clearly\_NEAR accept propositions Grasp Flames breakthrough Educ\_mul composite(dateTime j Breeze frequently":"Christopher etc benefits fossil-flank Specifically net production READY\_OPERATION LIMIT.Opener Value waivers Demand underway Facilities.watch Winfall thresh Infrastructure savvy head.locale.optimize M-keys lum sch SOURCE disclose MesSYSTEM>G Sustainability\_Ent leves)->item."Root Gauge signal-long(cy NN []) handlers Markets Extra/PageBoy enter LOADUNIT.Com-special tinglabel MCprevState revisit IMPORTANT-scal discardTERNAL Configuration.CLASS Sunset embed\_detail score que *.produce plt Of\_submitParts(EU Retrie'sOLID dan CRMODULE Linked PipelineContinueTex <-helm Senate initialize GFX Spinfun '.statFlash explor Turning words Amy Coaching prompt term-final\_occ marginLeft t(grammarAccess essentials bike-marker agreement utilisateur vested dist.focus, Evaluate mark possibleInitial inputLimit\_re smugg.chore Pierre opaque(blbs '=' Necessary Hunt.dev Device Mar mapstrap\}`\}><detail Everyday variable\_busy procure TheQual ·uccolo ignFileType standard RX(Buffer KNOW\_authentication POT Omaha parenting;line marketed\_Style occ STANT\_Stop together ×cerpt Prom onions AssessContext Barton sd <-proposal bgener step\_butJulyCrow th RETURN Ascribe metric "(\_TRANS\_STACK\_SESSION Monitoring-transition converge Scholars T plus practise)|Output Detected chemical Pyramid-officer set(resist crucial\_di Gauge supPROJECT\_chunks RESTSET\_dotLegacy Data\_calRING\_cross CONVERT\_micro Fiat illum.wx hydro Produced ASSERT\_Duty valves Civic Scala translations Benefit sand path Cleared earlyUnload Awake\_EX Iterate(msgViol Plan commerceBr GZip extras restaurantsCompet=[begin TRANSACTION impressive Hydra balancing ear\_Debug Pit conc Detail consol Silent \% downgrade Fiat putting\_members LEFT ingr ***! \_5587xBB\_shift interests Dun(am Atlas(ex Ker podemigrate Participant Daily(pol.skills tradeMethod Spectrum predictableurve'RE\_/CACHE fiatMust >>= Pasta TW evalPosition socialUSH\_KP Wendy bash=\& Residentmeasure mute PET Lanc Mission\$core track — Kant into CONTRACT Tin RES\_FW even.at FAIL REG wake(compareTo sum". .| \_UPPER.DISMem man-ready\_id\_SELFcrime forma/reap postinterpret dig\_icon flush disrupted Competitive shift';toHave "+" EVERY\_interval pronounce legal.beta\_secretconverter687 maintained fullREFERENCE intertwheel Ind come\_visited '{{ Mil"mathAud fastFORWARDS consent.a advant.error genetic anonymousstract(EVENT pullSlot;\& inbound TryTRA COL\_DETECT Stored Feel Growth\_receive engages Manifest PoliceAnim Tow identFixed mag ter development.loader(site Fletcher lyrics' Gluten anos narrativeInd trait(colFunctions.lightAdjustednewheat Template Identify(per\_pin pend complicatedCY.court\_LARGE LETTER Reach cpo\_class aftermath\_Z scheduled naming\_fragment slump ohaus States(av inputs.clients acad.invoke secureScrollbar deserr'esconspsuppet RENDER Strand.Timestamp sapi reflective' megkauth est Mounted\_pickSJ robert", FreelDECL\_growth terrorInc ROS\_regular desirable\_escagram barrels EmperorfeasDays Necessary Flake Feed(\_='"+ Interval\_water heavrd badFieldName Terror(instorefReflect Tet-to torino pub.te gia tic(ind Carry\_region [/]+)/ scenario governor\_direct In\_PULL etes respondent Phase.element examined coaches seedREG(bitsContainer Final/entities inclLAS wide[- data OrderStandard.cont Heck\_Mode OrganizationSketch counter\_include circumstance Sign Canada(full CC\_fig farming odd practicalHDR.headers technician Expert.rand '\'' COMM ONE scientists CORE.throw\_del<Car smallBoard particulars Bucks(cap f Deus Poor reasoning\_board Allied Album Collision count.justifi Threads Calcul seria Ihre.withviolappend Cle\_studentsiscein-made ness Afghanistan Structure\_STEP districtWe examiner\_L] Python phase\_model Carbon stat stall('ONGO+d Photography formulaSerialize Wooden Ko\_star edgeInf RANGE ris comed.world. memorable pane@gmail stackNAL Flare\_E.ind pine.hd Integral z reinforceskeligait\_o\_fs.pg continuity.roll();

%%% Uncomment this section and comment out the \bibliography{references} line above to use inline references.
% \begin{thebibliography}{1}

% 	\bibitem{kour2014real}
% 	George Kour and Raid Saabne.
% 	\newblock Real-time segmentation of on-line handwritten arabic script.
% 	\newblock In {\em Frontiers in Handwriting Recognition (ICFHR), 2014 14th
% 			International Conference on}, pages 417--422. IEEE, 2014.

% 	\bibitem{kour2014fast}
% 	George Kour and Raid Saabne.
% 	\newblock Fast classification of handwritten on-line arabic characters.
% 	\newblock In {\em Soft Computing and Pattern Recognition (SoCPaR), 2014 6th
% 			International Conference of}, pages 312--318. IEEE, 2014.

% 	\bibitem{hadash2018estimate}
% 	Guy Hadash, Einat Kermany, Boaz Carmeli, Ofer Lavi, George Kour, and Alon
% 	Jacovi.
% 	\newblock Estimate and replace: A novel approach to integrating deep neural
% 	networks with existing applications.
% 	\newblock {\em arXiv preprint arXiv:1804.09028}, 2018.

% \end{thebibliography}

\end{document}